\documentclass[numberedappendix,apj,iop]{emulateapj}
\usepackage{apjfonts}
\DeclareMathSymbol{v}{\mathord}{cmletters}{"76}
\usepackage{amsmath}
\usepackage{amssymb}
\usepackage{hyperref}

\newcommand{\useiop}{0}

\newcommand\figwidthfactor{0.8}

\usepackage{ifthen}
\usepackage[usenames,dvipsnames]{color}

\defcitealias{bz77}{BZ77}

\newcommand{\detg}{{\sqrt{-g}}}
\newcommand{\dF}{{^{^*}\!\!F}}

\newcommand{\astar}{a}

\newcommand{\OmegaH}{\Omega_{\rm H}}

\newcommand{\myfrac}[2]{{^{#1}\!/\hspace{-1pt}_{#2}}}

\newcommand{\gdet}{\sqrt{-g}}

\newcommand{\abs}[1]{\ensuremath{\left|#1\right|}}

\newcommand{\cut}[1]{\hbox{}}

\submitted{Accepted \today}

\shortauthors{A. Tchekhovskoy, R. Narayan, \& J.~C. McKinney}
\shorttitle{Black Hole Spin and the Radio Loud/Quiet Dichotomy of
  Active Galactic Nuclei}

\begin{document}
\label{firstpage}

\title{Black Hole Spin and the Radio Loud/Quiet Dichotomy of
  Active Galactic Nuclei}

\author{Alexander Tchekhovskoy,$^1$ Ramesh
  Narayan$^1$, Jonathan C. McKinney$^2$ }
\altaffiltext{1}{Institute for Theory and Computation,
  Harvard-Smithsonian Center for Astrophysics, 60 Garden Street,
  Cambridge, MA 02138, USA; atchekho@cfa.harvard.edu,
  rnarayan@cfa.harvard.edu} \altaffiltext{2}{Kavli Institute for
  Particle Astrophysics and Cosmology, Stanford University, P.O. Box
  20450, Stanford, CA 94309, USA; Chandra Fellow;
  jmckinne@stanford.edu}

\begin{abstract}

  Radio loud active galactic nuclei (AGN) are on average 1000 times
  brighter in the radio band compared to radio quiet AGN.
  We investigate whether this radio loud/quiet dichotomy can be due to
  differences in the spin of the central black holes that power
  the radio-emitting jets.  Using general
  relativistic magnetohydrodynamic simulations, we construct steady
  state axisymmetric numerical models for a wide range
  of black hole spins (dimensionless spin parameter $0.1\le \astar \le 0.9999$) and a
  variety of jet geometries.  We assume that the total magnetic flux
  through the black hole horizon at radius $r_{\rm H}(\astar)$ is held constant.
  If the black hole is surrounded by a thin accretion disk, we find that
  the total black hole power output depends
  approximately quadratically on the angular frequency of the hole,
  $P\propto\OmegaH^2\propto (\astar/r_{\rm H})^2$.  We conclude that,
  in this scenario, differences in
  the black hole spin can produce power variations of only a few tens at
  most.  However, if the disk is thick such that the jet subtends a narrow
  solid angle around the polar axis, then the power dependence
  becomes much steeper, $P\propto\OmegaH^4$ or even $\propto \OmegaH^6$.  Power variations of $1000$ are then possible
  for realistic black hole spin distributions.
  We derive an analytic solution that accurately
  reproduces the steeper scaling of jet power with $\OmegaH$,
  and we provide a numerical fitting formula that reproduces
  all our simulation results.
  We discuss other physical effects that might contribute to the
  observed radio loud/quiet dichotomy of AGN.

\end{abstract}



\section{Introduction}
\label{sec_intro}

The first active galactic nuclei (AGN) were discovered through radio
emission associated with their relativistic jets.  However, it soon
became clear that not all AGN\footnote{We use the generic term AGN to
refer to both luminous quasars and less luminous active nuclei such as
Seyferts, LINERs, etc.} have powerful radio jets; in fact, only about
10\% of quasars do.  The evidence for a dichotomy between radio loud
and radio quiet AGN has become stronger over the years, culminating in
the impressive demonstration by \citet*{ssl07} that two very distinct
populations of AGN are clearly visible when radio luminosities $L_R$
of AGN are plotted against optical luminosities $L_B$.  For a
given value of $L_B$, these authors show that $L_R$ of radio loud
AGN is $\sim10^3{-}10^4$ times greater than that of
radio quiet AGN.  Also, the two populations follow two
well-separated tracks on the plot.

The origin of the radio loud/quiet dichotomy has been much discussed
in the literature.  At Eddington ratios $\lambda = L_{\rm bol}/L_{\rm
Edd} \sim 10 L_B/L_{\rm Edd} > 0.01$, where $L_{\rm bol}$ is the
bolometric luminosity of the AGN and $L_{\rm Edd}$ is its Eddington
luminosity, a likely explanation for the dichotomy \citep{ho00,ssl07}
is that these systems accrete via a standard thin accretion disk.  Jet
production is then expected to be intermittent, as found to be the
case in black hole (BH) X-ray binaries (\citealt{fend04a}).  However, the
existence of two distinct populations for $\lambda<0.01$ is harder to
explain.  Even at these low luminosities, the radio loudness parameter
$R=L_{\rm 5\ GHz}/L_B$ of the radio loud population is at least a
factor of $10^3$ times larger than that of the radio quiet population.
However, at low values of $\lambda$, BH X-ray binaries typically are in a hard
spectral state associated with an advection-dominated accretion flow
(ADAF, \citealt{nm08}), and in this state, all BH X-ray binaries are radio loud
\citep{fend04a}.  Why then do AGN with similar values of
$\lambda$ have a radio loud/quiet dichotomy?

One possible explanation is that radio loud objects are driven by a
central BH with a large spin which produces a jet by the Blandford-Znajek
(BZ) mechanism \citep[hereafter, \citetalias{bz77}]{bz77}.  This is
referred to as the spin paradigm
\citep{blandford1990,wc95,blandford99}, which is in contrast to the
accretion paradigm which states that the BH mass and mass
accretion rate determine the jet power
\citep{blandfordrees74,blandfordrees92}.  These different paradigms
plausibly operate together to some degree \citep{bbr84,meier02}.  The
spin paradigm has been invoked to explain the observed correlation
between jet and accretion power in elliptical galaxies
\citep{allen2006} by combining an ADAF accretion model with the BZ
effect \citep{nemmen07,benson09}.  In terms of the dimensionless spin
parameter $\astar = J/GM^2$, where $M$ and $J$ are the mass and
angular momentum of the BH, it is found that one requires
$\astar\gtrsim 0.9$ to explain the correlation.  The dichotomy in the
power and spatial distribution of emission in Fanaroff-Riley classes 1
and 2 (FR 1 and FR 2) radio galaxies may also be explained by the spin
paradigm \citep{baum95}.

The radio loud/quiet dichotomy in AGN \citep{kellerman89,msl98,ivezic04}
has been explained in terms of an in situ trigger for relativistic jets
\citep{meier97}. 
It could also be explained by the differences in the evolutionary
stages at which we observe the objects \citep{blundell08} coupled with
the episodic activity of the AGN evidenced by the 
change or precession of jet orientation \citep{sb08,sbd08,shb10}.
However, another possibility is that the two AGN populations
have different merger and accretion histories which lead to different BH spins.
Recent observations show that, for $\lambda<0.01$, all radio loud
AGN reside in elliptical galaxies, whereas radio quiet AGN live mostly in
spirals \citep{ssl07}.
\citet*{volonteri07} explored a number of scenarios for the
formation of ellipticals and spirals and showed that it is plausible
for the nuclear BHs in spirals to have lower spins than those in ellipticals.
This suggests that the spin paradigm may explain the radio loud/quiet dichotomy.

The main problem with this explanation is that the difference
in radio loudness between the radio loud and radio quiet populations is a factor of $10^3$ \citep{ssl07}.
This is a strikingly large difference.
According to the accretion or spin paradigms,
relativistic jets are produced by magnetic outflows from either the inner region of the
disk or the spinning BH.  The power in the disk outflow is expected to
be proportional to the disk luminosity, which leaves no room for a
radio loud/quiet dichotomy, so we will ignore this
possibility\footnote{The BH spin also drives power into the disk causing a
more powerful disk outflow, but this still requires BH spin to introduce a dichotomy.}.
The power from the BH does depend on the spin parameter, and we will focus
on this\footnote{There are arguments to suggest that the luminosity of
  the disk outflow should be greater than that from the central
  spinning BH \citep{ga97,lop99}.  However these arguments assume low
  values of turbulent viscosity and weak magnetic fields near the BH,
  and also do not account for the effects of the general relativistic
  plunging region (see, e.g., \citealt{mck07a} for a discussion).
  \citet{mck05} finds that the luminosities of the jet and the
  disk are similar (however, see \citealt{nagataki09}).  
  For the purposes of this paper, we ignore the disk
  wind. 
}.
However, the analytical model of \citetalias{bz77}
indicates that the jet power $P$ varies only as $\astar^2$
for fixed magnetic flux threading the horizon.
If such a weak variation is to produce a difference of $10^3$ in radio
power\footnote{We assume that the radio luminosity is proportional to
  the jet power.}, we need $\astar$ in the two populations to differ by
a factor $\sim30$, which does not seem plausible given likely
merger histories \citep{hughesblandford03,gammie_bh_spin_evolution_2004}.
More plausible is a factor $\sim3$ difference in the median values of $\astar$ in the two
populations (e.g., see \citealt{volonteri07}), but this will produce
only a factor $\sim10$ difference in jet power, not $10^3$.

The \citetalias{bz77} scaling for power, $P\propto \astar^2$,
was derived in the limit $\astar\ll1$, for a razor-thin disk, assuming
the magnetic flux threading the BH is independent of $\astar$.
Recent analytical and numerical work by \citet{tn08} show that,
for a BH threaded by a split monopole magnetic field,
$P$ increases as $\astar^4$ at large values of $\astar$ when
higher-order corrections are included.
However, the analytical model worked out by these authors only gives a factor of two increase in
power above the \citetalias{bz77} result
even at $\astar=1$.
Their numerical simulations achieve a slightly steeper scaling,
but still only a factor of four above the \citetalias{bz77} result at $\astar=1$.
In any case, their work hints that a much steeper dependence of power on $\astar$ occurs as $\astar\to 1$.
Are there other effects that can introduce an even steeper dependence on $\astar$?

General relativistic magnetohydrodynamic simulations of accretion disks
by \citet{mck05} showed that for $\astar\gtrsim 0.5$
the jet power varies as steeply as the fourth power of the BH
angular rotation rate, i.e., $P\propto \OmegaH^4$, where $\OmegaH\propto \astar/r_{\rm H}$
and $r_{\rm H}$ is the radius of the horizon.
Compared to the scaling $P\propto\astar^4$, the scaling $P\propto \OmegaH^4$ introduces an additional factor of $16$
due to the division by $r_{\rm H}$, since $r_{\rm H}$ decreases from $2M$ to $M$
as $\astar$ varies from $0$ to $1$.
\citet{mck05} also finds that the power output of the entire BH has a shallower dependence on $\OmegaH$
compared to the power output of the jet, which subtends
a small solid angle above the disk and corona\footnote{\citeauthor{mck05}'s models
have an accretion disk with a disk+corona+wind of angular extent $H/R\sim 0.6$.
For $\astar\gtrsim 0.5$ the power per unit mass accretion rate scales as $\propto \OmegaH^5$
for the polar jet and $\propto \OmegaH^4$ for the entire horizon.
However, in these models, the mass accretion rate through the BH horizon
per unit fixed mass accretion rate at large radius scales as $1/\OmegaH$ for $\astar\gtrsim 0.5$, because of
the ejection of a massive wind (as also seen in \citet{hk06}).
Hence, expressed in terms of $\dot{M}$ at large radius, the polar jet power scales $\propto \OmegaH^4$ and the power output from the entire horizon scales $\propto \OmegaH^3$.}.
This suggests that changes in the solid angle subtended by the jet (via changes in the disk thickness)
could change the steepness of the power output as a function of $\astar$.

Since the scaling of $P$ with
BH spin is important for jet studies, we have explored this issue in
detail using general relativistic magnetohydrodynamic numerical simulations.
We consider a variety of geometries for the shape of the jet to see if we can come up with any
scenario in which jet power could change by a large factor for a
modest variation in $\astar$. We show that the most favorable scenario
is a BH surrounded by a thick accretion flow with an angular thickness
$H/R\sim1$.  We show that in this case the power output into a polar jet
has a steep dependence on the spin, $P\propto \OmegaH^4$, and that
the scaling steepens to $P\propto \OmegaH^6$ for even thicker disks.
Hence, we confirm the basic result found by \citet{mck05}
of a steep dependence of jet power on $\astar$ at high latitudes above the disk.
We suggest that this strong dependence may explain
the radio loud/quiet dichotomy in AGN.

Our numerical setup is described in \S\ref{sec:num}.  The results
for BHs with razor-thin disks are presented in \S\ref{sec:thindisk}, and
for jets from BHs with thick disks in
\S\ref{sec:thickdisk}.  We discuss the results in the context of the AGN radio
loud/quiet dichotomy in \S\ref{sec:discussion}, and conclude in
\S\ref{sec:conclusions}.  We work with Heaviside-Lorentzian units and
set $c=G=1$.

\section{Numerical Setup}
\label{sec:num}

\begin{figure*}
  \begin{center}
      \includegraphics[width=\textwidth]{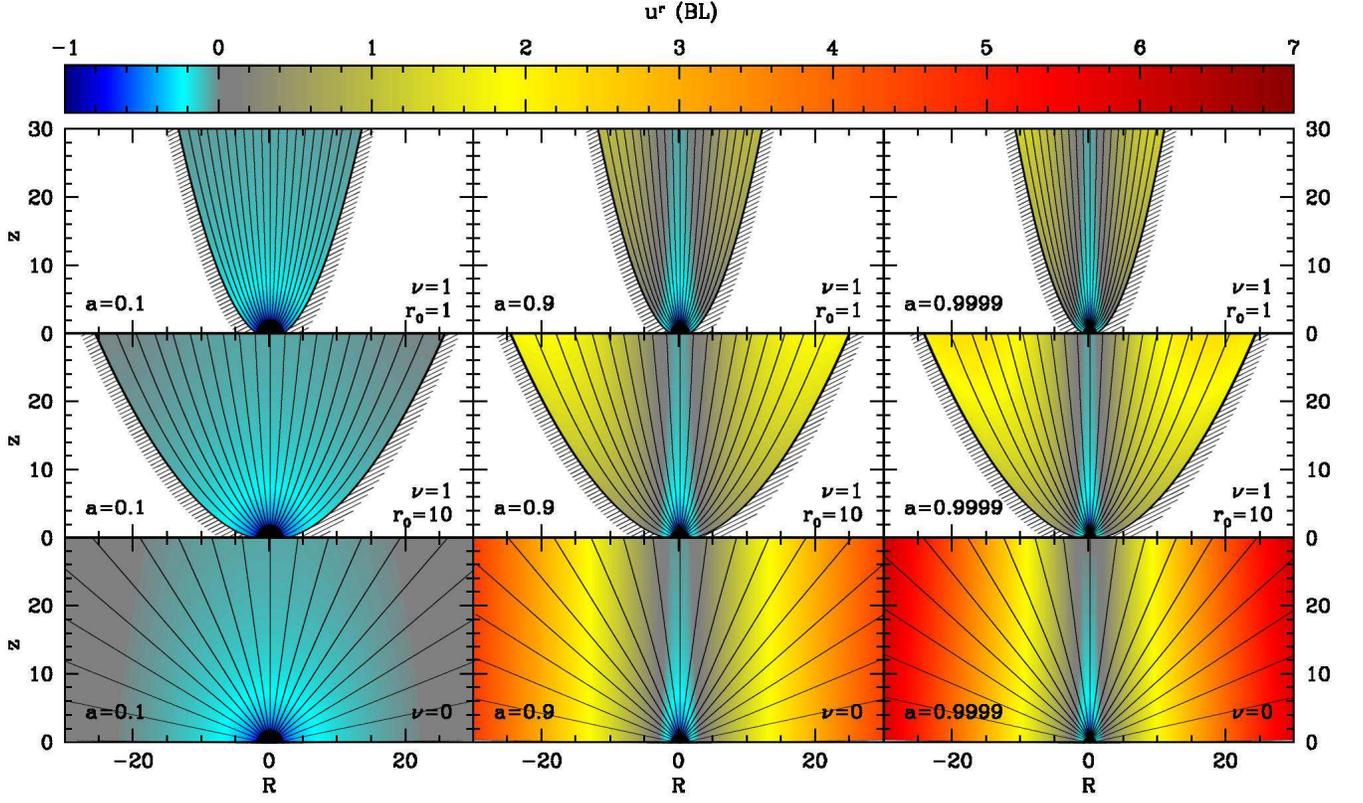}%
  \end{center}
  \caption{
    Meridional cuts through a selection of numerical jets in the force-free
    approximation, i.e., neglecting plasma inertia.  The
    color-coded radial four-velocity $u^r$ (as measured in
    Boyer-Lindquist coordinates, see legend) is shown overlaid with poloidal
    field lines (thin black lines which correspond to
    $\Psi^{1/2}=0,0.2,\dots,1$).  The jets are confined by collimating
    walls (thick black lines).  The left-most column of panels shows
    models with a low BH spin ($\astar=0.1$), the right-most column models with a
    fast spin ($\astar=0.9999$), and the middle column models with an
    in-between spin ($\astar=0.9$).  The top row of panels shows the most
    collimated paraboloidal models ($\nu=1, ~r_0=M$), the bottom row
    non-collimating monopolar models ($\nu=0$), and the middle row
    models that are monopole-like until $r\simeq{}r_0=10M$,
    beyond which they become paraboloidal ($\nu=1,r_0=10M$).}
  \label{fig1}
\end{figure*}

\begin{figure*}
  \begin{center}
      \includegraphics[width=\textwidth]{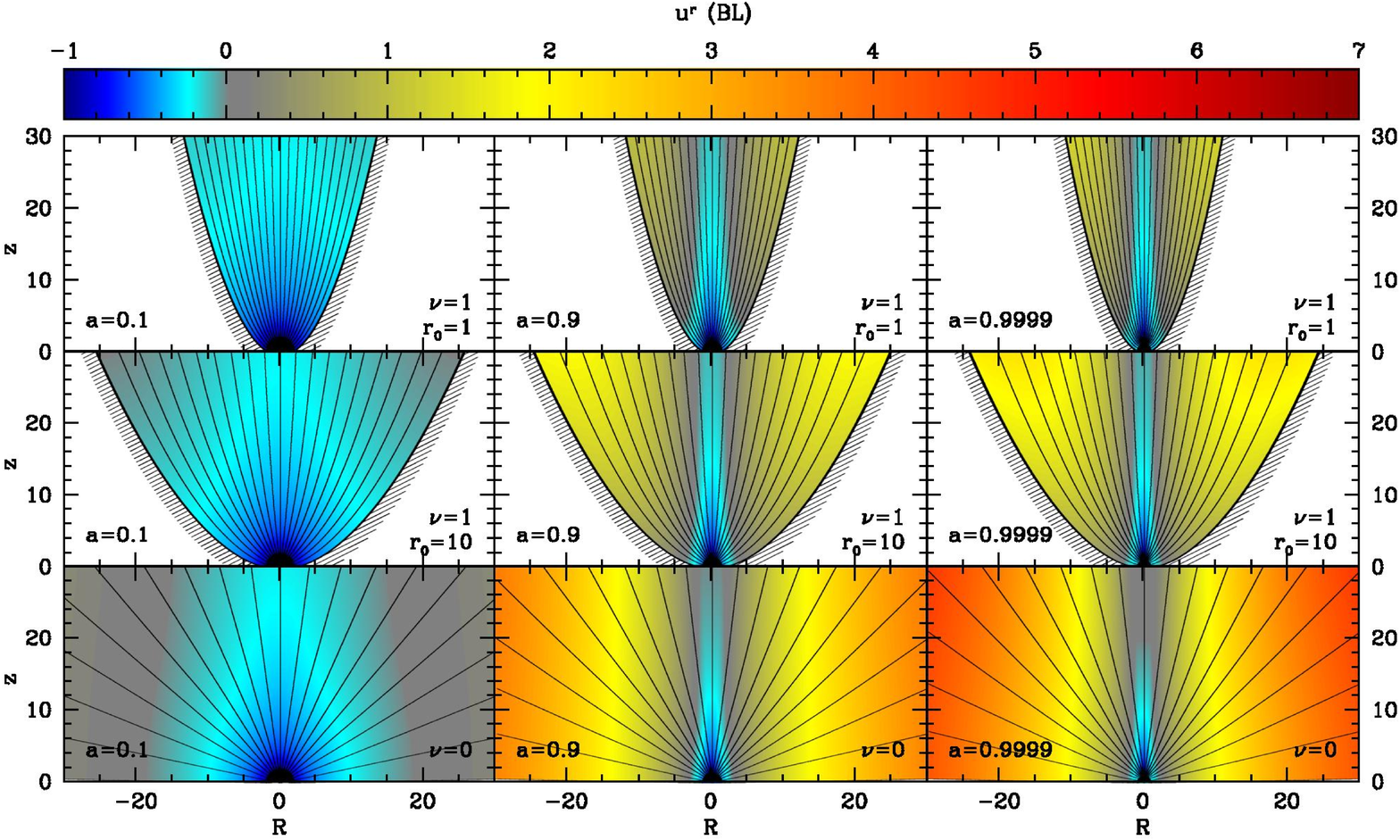}%
  \end{center}
  \caption{ Meridional cuts through a selection of numerical
    magnetohydrodynamic, i.e., mass-loaded, jets.  These may be directly
    compared to the force-free (i.e., neglecting plasma inertia) jet models shown in
    Figure~\ref{fig1}: the jet geometries are the same,
    but the jets here include mass-loading.
    While the field line shape
    changes little from Figure~\ref{fig1}, the introduction of
    mass-loading does lead to clear differences in the velocity.  (i)
    The inflow near the BH is faster in mass-loaded jets ($u^r$ is more
    negative in blue regions) due to the gravitational pull of the BH.
    (ii) The outflow from the BH in
    monopolar jets (the bottom set of panels) is slower than in the
    corresponding force-free simulations.  Nevertheless,
    the jet powers are nearly the same in the two sets of models.}
  \label{fig2}
\end{figure*}

It is known that a highly magnetized relativistic jet does not easily self-collimate.
For instance, the equilibrium field configuration
around an isolated spinning BH threaded by a magnetic field (sourced
by external currents) takes the form of a split monopole (\citetalias{bz77}).
Only extremely close to the polar axis is any evidence of self-collimation
present \citep{tch09}.
Therefore, in order to produce a jet which collimates most of the
energy output from the BH, it is necessary
to introduce an external confining medium.  The confining agent may be
the gas in an accretion disk, a corona, or a wind emerging from the
inner regions of the disk.  Ideally, one should numerically simulate
both the jet and the confining medium, but this is numerically very
challenging.  Instead, we follow the more usual approach (e.g.,
\citealt{kom07,tch09b}) of introducing a rigid axisymmetric wall with a prescribed
shape and requiring the jet to lie inside the wall.  The shape of the
wall is set by two parameters:
\begin{enumerate}
\item An index $\nu$ which sets the asymptotic poloidal field line
shape, as described below. This parameter ranges from $\nu=0$, which
corresponds to a monopole field geometry, to $\nu=1$, which
corresponds to a paraboloidal jet.  In a real system, $\nu$ would be
set by the radial pressure profile of the confining medium.  Plausible
values are in the range $\nu\sim0.5{-}1$ \citep{tch08}.
\item A transition radius $r_0$ which is defined such that for
\hbox{$r\lesssim{}r_0$} the jet is monopolar and for \hbox{$r\gtrsim{}r_0$} it
switches to the shape prescribed by $\nu$. The parameter $r_0$ allows
us to consider situations in which confinement operates only beyond a
certain distance from the BH.
\end{enumerate}

In terms of these two parameters, the wall has the following shape in
polar $(r,\theta)$ coordinates in the Boyer-Lindquist frame:
\begin{equation}
  \label{eq:theta_asym}
  1-\cos\theta = \left(\frac{r+r_0}{r_{\rm H}+r_0}\right)^{-\nu},
  \qquad r_{\rm H}=M\left[1+(1-\astar^2)^{1/2}\right],
\end{equation}
where $r_{\rm H}$ is the radius of the BH horizon.  For $r\ll{}r_0$,
$(1-\cos\theta)\ll1$, so $\theta\approx\pi/2$, i.e., the wall lies
along the equatorial plane, as for a split monopole.  For $r\gg(r_{\rm
H},r_0)$, $\theta\propto{}r^{-\nu/2}$, which corresponds to a
generalized paraboloid.  Note that, in all these models, the wall
meets the horizon at the equator.  In effect, this means we assume a
razor-thin disk which subtends zero solid angle at the BH.  In \S3.2
we discuss the case of geometrically thick disks.

\begin{figure*}
  \begin{center}
      \includegraphics[width=0.9\textwidth]{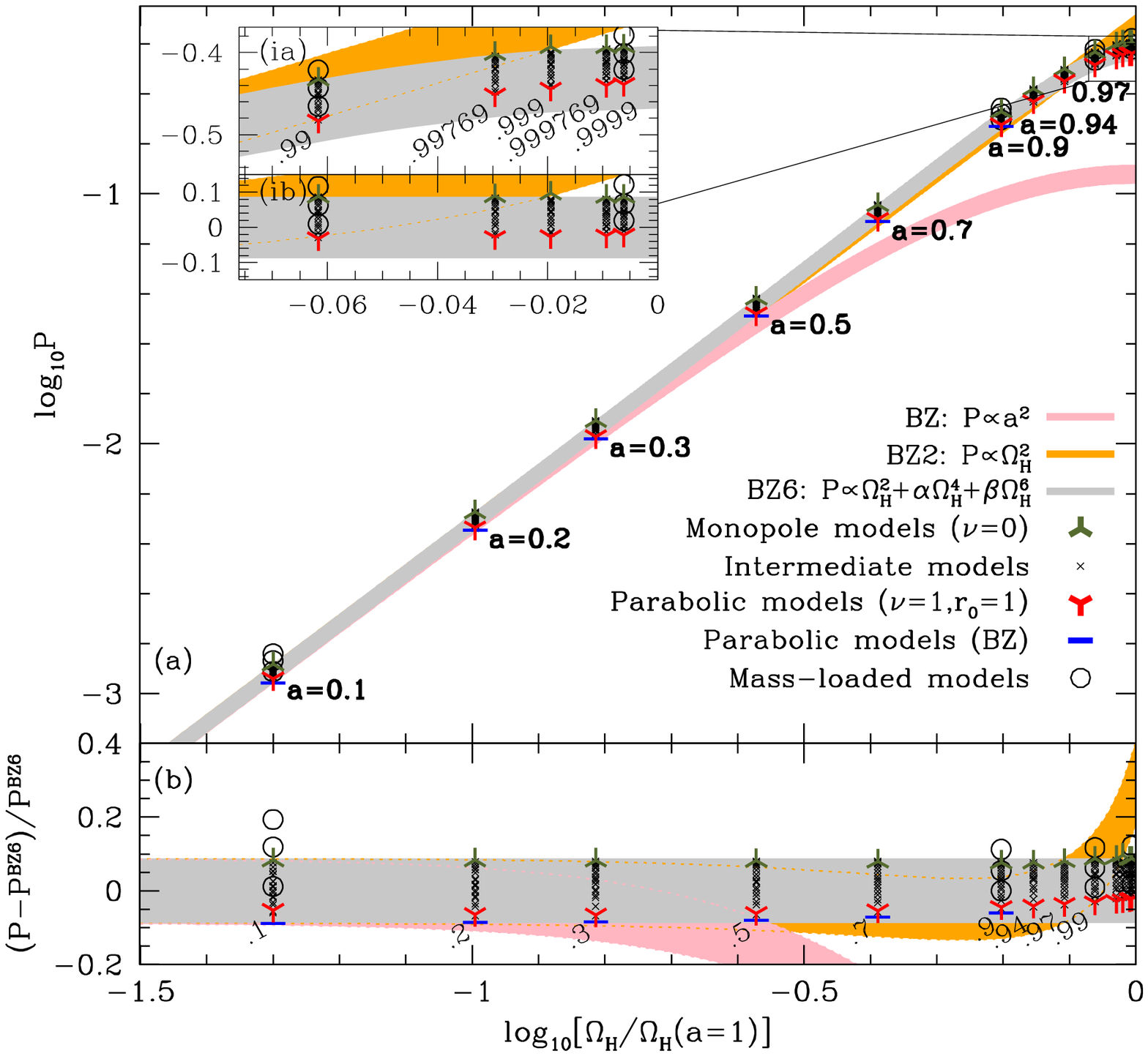}%
  \end{center}
  \caption{Jet power output of various models as a function of
dimensionless BH horizon spin
    frequency $\omega_{\rm H}=\OmegaH/\OmegaH(a=1)$ (equation~\ref{eq:aeff}). Different models
    are shown with different symbols (see legend). For reference, the following 
    values of BH spin, $a=\{0.2,0.5,0.9,1\}$, correspond to approximately
    the following values of $\omega_{\rm H}$: $\{0.1,0.27,0.63,1\}$.   
    [Upper panel (a)]:
    Logarithm of jet power. [Lower panel (b)]: Fractional
    deviation of the numerical jet power from the BZ6
    formula~\eqref{eq:bzbf}. [Insets (ia) and (ib)] Blow-ups
    of panels (a) and (b) for high values of BH spin. The values of the BH
    spin $\astar$ are shown as labels next to data points.  The three colored
    stripes (see legend) correspond to the original \citetalias{bz77} scaling (equation~\ref{eq:bzori}), the BZ2 scaling (equation~\ref{eq:bzaeff}), and
    the BZ6 scaling (equation~\ref{eq:bzbf}).
    The second-order BZ2 scaling~\eqref{eq:bzaeff} follows the numerical results very well except
    for values of $\astar$ close to unity.
    The BZ6 scaling~\eqref{eq:bzbf} matches the numerical data well
    at all values of $\astar$. For each $\astar$, the monopolar model ($\nu=0$)
    has the maximum power and the paraboloidal model has the least power.
    However, the difference is only $\sim20\%$}
  \label{fig3}
\end{figure*}

Having picked the shape of the wall, we choose the poloidal flux
function of the initial magnetic field configuration to be
\begin{equation}
  \label{eq:psi}
  \Psi=\left(\frac{r+r_0}{r_{\rm H}+r_0}\right)^\nu(1-\cos\theta).
\end{equation}
Note that $\Psi$ is conserved along each field line, and
$\Phi=2\pi\Psi(r,\theta)$ is the poloidal magnetic flux through a toroidal
ring at $(r,\theta)$.
By construction, equation \eqref{eq:psi}
corresponds to a total flux of $\Phi_{\rm tot}=2\pi$ in the jet.
In this model $\Phi_{\rm tot}$ does not depend upon spin,
so the amount of magnetic flux threading the BH is fixed for different spins.
The outermost field line, defined by $\Psi=1$, follows the
shape of the wall.  This particular field line is always located at the
wall because of our boundary conditions. Interior field lines,
however, are free to move once the simulation begins and do experience
minor shifts in the poloidal direction.  The initial magnetic
field has no toroidal component.

There are no known exact solutions for the magnetosphere of a spinning
BH.  However, the poloidal field configurations given in
equation~\eqref{eq:psi} are sufficiently close to the true solutions
that their initial relaxation when the simulation starts is rather
mild.  Only two linearly independent analytic solutions have been obtained for a
non-spinning BH: one corresponds to a monopolar field geometry and is
given by equation~\eqref{eq:psi} with $\nu=0$, while the other is the
following \citetalias{bz77} solution for a paraboloidal field,
\begin{equation}
  \label{eq:bzpsipara}
  \Psi=\frac{(r/r_{\rm H}-1)(1-\cos\theta)-(1+\cos\theta)\log(1+\cos\theta)}{2\log2}+1.
\end{equation}
Note that the split-monopole solution applies to the entire space exterior to the horizon,
whereas the paraboloidal solution only applies to the field lines attached to
the BH (see \citealt{mck07b} for a numerical paraboloidal solution that applies
to the whole space). Any linear combination of the monopolar and
paraboloidal solutions is also a solution \citep[see, e.g.,][]{bes09}.
The \citetalias{bz77} paraboloidal solution~\eqref{eq:bzpsipara} is very similar to the approximate
solution~\eqref{eq:psi} for the case $\nu=1$, and this is our reason
for focusing on the simpler model~\eqref{eq:psi}, with $\nu$ varying over the range 0 to 1.
For completeness, we have also run simulations using the \citetalias{bz77}
field geometry~\eqref{eq:bzpsipara} to initialize the calculations
(along with the appropriate choice of the wall shape, obtained by
setting $\Psi=1$ in this equation).

We performed the simulations using the general relativistic (GR) MHD
code HARM \citep{gam03,mck04,mck06jf,mck06ffcode,nob06}
using Kerr-Schild coordinates in
the Kerr metric; the code includes a number of recent improvements
(\citealt{mm07,tch07,tch08,tch09}). Most of the simulations were done in the
force-free approximation, which assumes that the plasma is infinitely
magnetized and has negligible inertia.  Within this approximation, the
problem is fully defined by specifying just the BH spin and the shape of the wall.  Figure~\ref{fig1} shows
results from several representative simulations.  

Real relativistic
jets are of course not perfectly force-free; in fact, they are
expected to deviate substantially from this approximation at large
distances from the BH.  However, all relativistic MHD jets that have
$\gamma\gg1$ are highly
magnetized near the BH, and here they are expected to be well
represented by force-free solutions \citepalias{bz77}.  Moreover, the power
that a relativistic jet carries is determined entirely by the initial
force-free zone.  Therefore, we expect numerical results on the jet
power from a force-free simulation to agree very well with the power
for an MHD jet with inertia.

To verify this expectation, we have repeated some of our force-free
simulations in the MHD limit, in which the jet is mass-loaded with a
finite amount of plasma (details given below).  Figure~\ref{fig2}
shows some results.  As expected, we find that the asymptotic Lorentz
factor $\gamma$ of the jet does depend on mass-loading: a force-free
jet accelerates without limit \citep{tch08}, whereas an MHD jet
asymptotes to a finite $\gamma$ which is determined by the initial
magnetization of the jet.  However, we find that the jet {\it power},
the primary quantity of interest to us in this paper,
is not sensitive to mass-loading so long as the jet is relativistic,
i.e., so long as the jet is initially force-free near the BH.

MHD jets are more complicated and require more parameters to be specified compared
to force-free jets.  In particular, the results depend on the details of mass-loading
at the base of the jet.
Highly magnetized jets accelerate because the magnetic energy flux, which dominates
the energy budget at the base of the jet, is converted to kinetic
energy flux of the plasma as the jet flows out.  The ratio of magnetic
to kinetic energy flux at the base of the jet determines the
asymptotic Lorentz factor \citep{begelman_asymptotic_1994,kom09,tch09,tch09b}.  Observations
suggest a characteristic value for the Lorentz factor of AGN jets
$\gamma_{\rm AGN}\sim25$ \citep{jorstad_agn_jet_2005}.  We choose the
following simple prescription for the mass-loading of our numerical
MHD jets to roughly match this value.  We impose a floor on the
co-moving rest-mass density of the jet $\rho_{\rm floor}$ such that
whenever the density falls below this value, we add mass in the
co-moving frame of the plasma.  This simple floor model is a
convenient way of numerically representing more complicated (and
poorly understood) processes that are responsible for the mass-loading
of jets in AGN.  The value of $\rho_{\rm floor}$ is selected such that
the rest mass energy density $\rho_{\rm floor}c^2$ is equal to a
fraction $1/\gamma_{\rm AGN}$ of the co-moving magnetic energy density $\epsilon_m$:
$\rho_{\rm floor} c^2 = \epsilon_{\rm m} / \gamma_{\rm AGN}$.  Thus,
our floor model ensures that the ratio of $\epsilon_m$ to $\rho_{\rm
floor}c^2$ does not exceed $\gamma_{\rm AGN}$, thereby making sure
that the maximum Lorentz factor of our jets is close to the required
value.  We note that while our procedure is Lorentz invariant, it
might not correspond to a physical process that operates in AGN, e.g.,
photon annihilation from the accretion disk~\citep{phi83}.

The code uses internal coordinates $(x_1,x_2)$ that are uniformly
sampled with $512\times128$ grid cells.  The internal coordinates are
mapped to the physical coordinates $(r,\theta)$ via $r/M=R_0+\exp(x_1)$
and $x_2={\rm{}sign}(\Psi)\abs{\Psi}^{1/2}$. The computational domain
extends radially from the inner boundary at $r_{\rm{}in}=0.6M+0.4r_{\rm H}$
to the outer boundary at $r_{\rm{}out}=10^4M$.  We apply absorbing
(outflow) boundary conditions at each of these boundaries.  In the
$\theta$-direction, the computational domain extends from the polar
axis at $x_2=0$, where we use the usual polar boundary conditions, to
the jet boundary $x_2=1$, where we place the wall.  At the wall, we
outflow (copy) the components of velocity and magnetic field that are
parallel to the wall, and mirror the perpendicular components
\citep{tch09}. For a given BH spin and grid resolution, we choose the
value of $R_0$ such that there are $7$ to $10$ grid cells between
$r_{\rm{}in}$ and $r_{\rm H}$. This ensures that the inner radial grid
boundary $r=r_{\rm{}in}$ is causally disconnected from the region
outside the BH horizon.

At time $t=0$ we initialize the simulation with a purely poloidal
field configuration as described in equation~\eqref{eq:psi} (or
equation~\ref{eq:bzpsipara} in the case of the \citetalias{bz77} paraboloidal model) and
we run the simulation until
$t_{\rm{}f}={\rm{}max}(100M,20/\OmegaH,10r_0)$.  Because of the
dragging of frames by the spinning BH, the magnetic field develops a
toroidal component which propagates out along field lines at nearly the speed
of light.  Behind this outgoing wave, the solution settles down to a
steady state and we study the properties of this steady solution.  We
have verified by running selected simulations for $10$ times longer
than our fiducial $t_{\rm{}f}$ that the near-BH regions of our
numerical solutions have truly reached steady state.

\section{Results}
\label{sec:results}

\subsection{Power Output of Black Holes with Razor-thin Disks}
\label{sec:thindisk}

As explained below equation~\eqref{eq:theta_asym}, all our jet models
correspond to the case of a razor-thin disk.  We describe here the
results we obtain for these models.

\citetalias{bz77} showed that the luminosity of a force-free jet from
a slowly spinning BH ($\astar\ll1$), embedded in a regular
magnetic field with a fixed total flux (sourced by toroidal currents
in a razor-thin disk), is proportional to the square of the BH spin and the square of
the magnetic field strength at the horizon.  If we include the length
scale of the horizon $2M$ to obtain the correct dimensions, we may
write (the choice of the numerical prefactor will become clear below)
\begin{equation}
  \label{eq:bzori}
  P^{\rm{}BZ}(\astar)=k \Phi_{\rm{}tot}^2\frac{\astar^2}{16M^2},
\end{equation}
where $\Phi_{\rm tot}\propto{}B{}M^2$ is the total poloidal magnetic
flux in the jet, and $k$ is a constant which depends on the field
geometry, e.g., $k=k_{\rm mono}=(6\pi)^{-1}\approx0.054$ for a monopolar field
($\nu=0$) and $k=k_{\rm para}\approx0.044$ for the paraboloidal \citetalias{bz77}
geometry (equation~\ref{eq:bzpsipara}).  We refer to
equation~\eqref{eq:bzori} as the original \citetalias{bz77} scaling.

Because equation (\ref{eq:bzori}) was derived in the limit $\astar\ll1$, it
can be extended to larger values of $\astar$ in several ways.  In
particular, we could replace the length scale $2M$ by the horizon scale
$r_{\rm H}$, where $r_{\rm H}$ is defined in equation~(\ref{eq:theta_asym}).  In
fact, this is a more natural scaling since the angular frequency of
the BH,
\begin{equation}
  \label{eq:aeff}
  \OmegaH(\astar)=\frac{\astar}{2r_{\rm H}(\astar)},
\end{equation}
clearly plays an important role in determining
the power in the jet at the horizon (\citetalias{bz77}; \citealt{mck04}).
\citetalias{bz77} found that,
for a fixed field strength at the BH horizon, the power in the outflow obeys
(\citetalias{bz77}; \citealt{mck04}):
\begin{equation}
  \label{eq:bhpower}
  P \propto \Omega(\OmegaH-\Omega)\bigr|_{r=r_{\rm
      H}} \propto  \OmegaH^2 ,
\end{equation}
where, based on dimensional argument, the field line
angular frequency, $\Omega$, is proportional to $\OmegaH$ (see
Appendix~\ref{app:bz2}).
Numerical simulations by~\citet{kom01}
showed that the \citetalias{bz77} effect achieves maximum efficiency
when $\Omega\approx \OmegaH/2$.  The result was found to be true for $\astar=\{0.1,0.5,0.9\}$,
demonstrating
that this result is valid even in the non-linear regime.
Now, replacing the length scale $2M$ with
the horizon scale $r_{\rm H}$ in the expression for power
\eqref{eq:bzori}, we obtain:
\begin{equation}
  \label{eq:bzaeff}
  P^{\rm{}BZ2}(\OmegaH){}=k\Phi_{\rm{}tot}^2\OmegaH^2.
\end{equation}
In Appendix~\ref{app:bz2} we derive this formula analytically from
first principles.
The scaling of jet power~\eqref{eq:bzaeff} was confirmed
in the numerical simulations by
Krasnopolsky (private communication).
In general, $k$ is a constant factor whose value depends on the field
geometry near the BH.  For a slowly spinning BH,
$\astar\approx4\OmegaH M$, and equation~\eqref{eq:bzaeff} reduces to
the standard \citetalias{bz77} scaling which was derived in the limit
$\astar\ll1$.  However, as expected from the above discussion and as
we will confirm shortly, equation~\eqref{eq:bzaeff} is a better
approximation for higher spins and is quite accurate up to
$\astar\approx0.95$, beyond which it requires a modest correction.  We
refer to equation~\eqref{eq:bzaeff} as the modified
\emph{second-order} \citetalias{bz77} scaling, or simply the BZ2
scaling.  We classify the order of a scaling by the maximum power of
$\OmegaH$ up to which the scaling maintains its accuracy.  As we will
see below, for large spins $a\simeq1$, scalings higher than the $2$nd
order are required to obtain good agreement with the numerical
results.

\citet{tn08} found $4$th order corrections to the
BH power output by performing the expansion in powers of BH spin
$a$.  As we have argued, a more accurate expansion is
in powers of the BH rotational frequency $\OmegaH$ (equation~\ref{eq:aeff}).
Recast in powers of $\OmegaH$, the \citet{tn08} expansion becomes:
\begin{equation}
  \label{eq:bz4}
  P^{\rm BZ4}(\OmegaH)\approx k\Phi_{\rm{}tot}^2(\OmegaH^2+\alpha\OmegaH^4),
\end{equation}
where the $4$th-order coefficient $\alpha = 8 \left(67-6
  \pi^2\right)/45\approx1.38$. We analytically derive this formula
from first principles in Appendix~\ref{app:bz4}.
Note that the expansion only contains
even powers of $\OmegaH$ since the power is independent of the sense
of BH rotation.  As we will see, this $4$th order
correction agrees well with the numerical results and is an
improvement over the $2$nd order formula~\eqref{eq:bzaeff}. However,
at high spins, $a\gtrsim0.99$, even the  formula~\eqref{eq:bz4}
becomes inaccurate.  Below we present a more accurate $6$th order formula
(see equation~\ref{eq:bzbf}).

We have performed numerical simulations of force-free jets confined by
a rigid wall (as described in \S\ref{sec:num}) for a wide range of
field geometries and BH spins.  We numerically explored all possible
combinations of
$\nu=\{0,\linebreak[0]0.25,\linebreak[0]0.5,\linebreak[0]0.75,\linebreak[0]1\}$,
$r_0=M\times\{0,\linebreak[0]1,\linebreak[0]5,\linebreak[0]10,\linebreak[0]100\}$
and
$\astar=\{0.1,\linebreak[0]0.2,\linebreak[0]0.3,\linebreak[0]0.5,\linebreak[0]0.7,\linebreak[0]0.9,\linebreak[0]0.94,\linebreak[0]0.97,\linebreak[0]0.99,\linebreak[0]0.99769,0.999,\linebreak[0]0.999769,\linebreak[0]0.9999\}$.
We also performed simulations in which we started the field with the
paraboloidal \citetalias{bz77} geometry~\eqref{eq:bzpsipara}.\footnote{We
note that this field geometry and the paraboloidal geometry with
$\nu=1,r_0=0$ are inherently difficult to study numerically: the wall
in these models makes such a small angle with the surface of the BH
horizon that
there exists no physical solution for the velocity in the immediate
vicinity of the wall.  We have obtained numerical solutions
corresponding to the paraboloidal \citetalias{bz77} model only for
$\astar\le0.9$ and the paraboloidal $\nu=1$ model only for $r_0\ge1$.}
In addition to force-free simulations, which neglect plasma inertia,
we have also performed MHD (mass-loaded)
simulations for selected field geometries and spins:
$\nu=\{0,\linebreak[0]1\}$,
$r_0=M\times\{0,\linebreak[0]1,\linebreak[0]10\}$ and
$\astar=\{0.1,\linebreak[0]0.9,\linebreak[0]0.99,\linebreak[0]0.9999\}$.

Figures~\ref{fig1} and \ref{fig2} illustrate the effect of the shape
parameter $\nu$ and the transition radius $r_0$ on the jet
geometry. The larger the value of $\nu$, the more collimated is the
jet.  The larger the value of $r_0$, the more monopolar-like is the
jet geometry near the BH. The shape of the poloidal field lines weakly
depends on the BH spin: careful examination of the figures reveals
that the field lines tend to come closer to the jet axis for faster
spins, and this effect is largest near the BH horizon.  The
magnetosphere clearly divides into an outflow region ($u^r>0$) and an
inflow region ($u^r<0$) separated by a stagnation surface at which the
radial velocity vanishes.  This is similar to what is seen in
simulations of magnetized turbulent tori around spinning BHs
\citep{mck04,mck06jf,mb09}.

Figure~\ref{fig3} shows the numerically measured power output of all
our models as a function of the BH horizon frequency $\OmegaH$
(defined in eq~\ref{eq:aeff}).  The colored stripes correspond to the three
scalings: BZ (equation~\ref{eq:bzori}), BZ2 (equation~\ref{eq:bzaeff}), and BZ6 (equation~\ref{eq:bzbf}).  For a given
BH spin, a monopolar field geometry ($\nu=0$) produces a more powerful
jet since it has a larger value of the pre-factor
$k_{\rm{}mono}\approx 0.054$ compared to the paraboloidal \citetalias{bz77}
geometry (equation~\ref{eq:bzpsipara}, $k_{\rm{}para}\approx0.044$,
\citetalias{bz77}).  This difference in $k$ determines the width of
the colored stripes in Fig.~\ref{fig3}.  The model given in
equation~\eqref{eq:psi} with $\nu=1$ is close to the \citetalias{bz77} paraboloidal
solution and has nearly the same power.  Models with intermediate
values of $\nu$ ($0<\nu<1$) or with non-zero values of $r_0$ have
power output intermediate between the two limiting models and form the
vertical clusters of numerical points at each $\astar$ in
Figure~\ref{fig3}.  Independent of the value of $\nu$, we find that
jets with $r_0$ much larger than the outer ``light
cylinder''\footnote{By the ``light cylinder'' we mean the
  Alfv\'en surface.} radius
$\simeq1/\OmegaH$ have luminosities very similar to that of a monopolar
jet.\footnote{This highlights the fact that jet power output is set by
the field line shape {\it close} to the BH, i.e., inside the light
cylinder.  It suggests that communication along the jet is
maintained by Alfv\'en waves (rather than fast waves), so that the
outer light cylinder, which acts as a sonic surface for Alfv\'en
waves, prevents signals propagating back to the BH from further out.
As a result, the shape of the wall or the properties of the confining
medium outside the light cylinder have no influence on the power
output in the jet.\label{ftn:locality}}

As expected, for small BH spins, $\astar\lesssim0.3$, both the
original \citetalias{bz77} scaling (equation~\ref{eq:bzori}) and the second
order BZ2 scaling (equation~\ref{eq:bzaeff}) agree well with the numerical
results.  As we go to higher spins, the BZ2
scaling~\eqref{eq:bzaeff} continues to follow the simulation data
points accurately, while the original \citetalias{bz77}
scaling~\eqref{eq:bzori} under-predicts the power (e.g., by a factor
$\approx3$ at $\astar=0.99$).

Careful examination of Fig.~\ref{fig3}a and especially of the inset
Fig.~\ref{fig3}ia, which shows a blowup of the $a\to1$ region of the plot,
reveals a flattening in the numerical data points at spin values
$\astar\gtrsim0.95$: the BZ2 formula \eqref{eq:bzaeff} over-predicts
the jet luminosity by about $25\%$ as $\astar\to1$ (in agreement with
R.~Krasnopolsky, private communication).  It is not clear that this
corner of parameter space is particularly relevant for astrophysics,
nor is the effect very large.  Nevertheless, for completeness we note
that the flattening of the jet power can be well-modeled by including
higher-order corrections to the BZ2 formula~\eqref{eq:bzaeff}, e.g.,
by the BZ6 formula which we derive in Appendix~\ref{app:bz6}.
We give here a simplified version of this formula:
\begin{equation}
  \label{eq:bzbf}
  P^{\rm{}BZ6}(\OmegaH)\approx k\Phi_{\rm{}tot}^2(\OmegaH^2+\alpha\OmegaH^4+\beta\OmegaH^6),
\end{equation}
where the value of $\alpha$ is
determined analytically, $\alpha \approx 1.38$ (same as in the BZ4 expansion,
equation \ref{eq:bz4}) and $\beta$ is found numerically
by least-square fitting equation~\eqref{eq:bzbf} to the full BZ6 analytic
formula derived in Appendix~\ref{app:bz6}: $\beta \approx -9.2$.
The gray stripe in Figure~\ref{fig3} compares this formula to our
numerical results for the full range of models.  Higher order
corrections have no effect at small spin values (the gray and light
red stripes lie on top of each other) but do a good job of reproducing
the flattening in the jet luminosity at spin values
$\astar\gtrsim{}0.95$ and the slight but systematic increase in the
power output of the numerical jets above the light red stripe at
$\astar\lesssim0.9$ (this increase is especially apparent in
Figure~\ref{fig3}b).  In anticipation of future discussion, it is
useful to express the power at low spin in terms of the maximum
achievable power at $\astar=1$:
\begin{equation}
  \label{eq:power_low_a}
  P(\astar)\simeq 0.32 \astar^2 P(\astar=1), \quad \astar\lesssim0.3.
\end{equation}

We now look into the origin of the differences in the power outputs of
the various model jets, as well as of the numerical trends discussed
above.  We focus on two limiting cases: monopolar jet ($\nu=0$) and
paraboloidal jet ($\nu=1$, $r_0=1$).

First, let us recast the power output of the jet in a convenient
form.
In a stationary axisymmetric force-free flow, several quantities are
conserved along poloidal field lines (defined by $\Psi={\rm{}const}$).
Two of these are the field line angular velocity $\Omega(\Psi)$ and
the enclosed poloidal current $I(\Psi)$ \citep{tch08}.  The power
output of a force-free jet may be written as the integral of the
outward Poynting flux $S^r\equiv -\Omega B^r B_\varphi$
over a spherical jet cross-section\footnote{Here the GR notation
is simplified and appears like the non-GR expressions (apart from some sign conventions)
by using the notational conventions in appendix B of \citet{mckinney2005} and in \citet{mck06ffcode}.
In this notation, $B^i\equiv\dF^{it}$, $B_i=\dF_{it}$, $E_i=F_{it}/\detg$,
$E^i=F^{ti}\detg$, and $\Omega\equiv -E_\theta/B^r$, where $F$ is the faraday tensor, $\dF$
is the Maxwell tensor, and $\detg$ is the square root
of minus the determinant of the metric.
Horizon surface area elements are given by $dA = \detg d\theta d\phi$.}.
In this notation, the lower component of the toroidal magnetic field is up to a
numerical factor the enclosed
poloidal current, $-2\pi B_\varphi\equiv I(\Psi)$.  Using this
notation, which is very similar in appearance and meaning to the
usual special relativistic notation, we obtain the total
power output of the BH by integrating over the surface of the BH (see also \citetalias{bz77}):
\begin{equation}
  \label{eq:jetpower}
  P=\iint{}S^r\,dA
  =2\int_0^{\pi/2}\Omega B^r I \,dA
  =2\int_0^1\Omega(\Psi)I(\Psi)\,d\Psi,
\end{equation}
where the field strength $B^r$ times the area element $dA$
gives the magnetic flux through that area, $d\Phi = 2\pi d\Psi=
B^r\,dA$, and the numerical factor of $2$ accounts for the two
hemispheres of the BH.

\begin{figure}
  \begin{center}
      \includegraphics[width=\figwidthfactor\columnwidth]{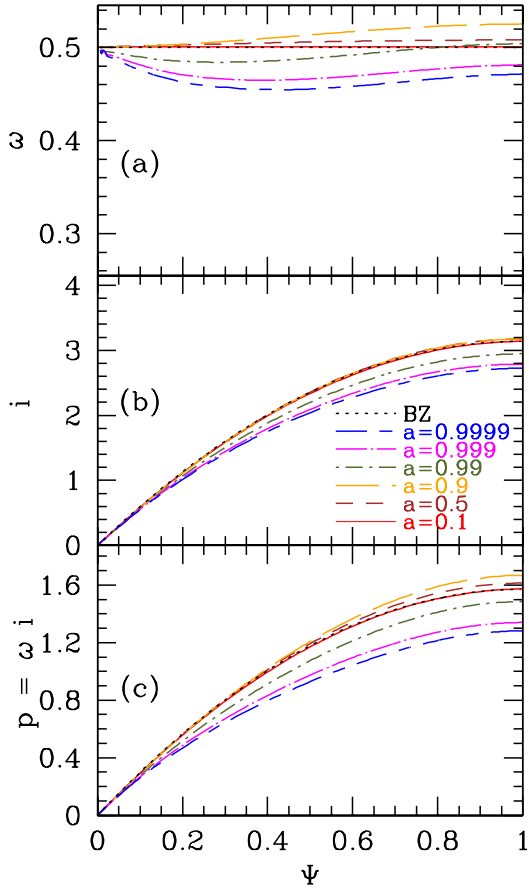}%
  \end{center}
  \caption{
    Angular dependence of various quantities
    in a monopolar jet ($\nu=0$) as a
    function of the poloidal flux function $\Psi$.
    The different curves in each panel correspond to different
    values of the BH spin (see legend).  From top to
    bottom the panels show the normalized field angular velocity
    $\omega=\Omega/\OmegaH$, the normalized enclosed poloidal current
    $i=I/\OmegaH$, and the normalized jet luminosity $p=\omega i =
    \Omega I/\OmegaH^2 = P/\OmegaH^2$.  The analytic \citetalias{bz77} solution, shown with
    dotted lines, provides an excellent description of the numerical
    results for all spin values $\astar\lesssim0.99$.  Beyond this value of $\astar$,
    the quantities $\omega$, $i$, and $p$ all become noticeably
    smaller than the analytic solution.  This trend is removed
    by the BZ6 solution~\eqref{eq:bzbf}, as shown in Figure~\ref{fig3}.}
  \label{fig4}
\end{figure}

Figure~\ref{fig4} shows for a monopolar jet the angular profiles of
angular velocity $\omega=\Omega/\OmegaH$, enclosed poloidal current
$i=I/\OmegaH$, and power output $p=P/\OmegaH^2$.
The particular scalings by $\OmegaH$ have been selected based on
equation~\eqref{eq:bzaeff} so as to remove any obvious trends as a
function of spin.  This allows us to compare models with different
spins on the same scale.  At low spin, $\astar\lesssim0.1$, we have
excellent agreement between the numerical models and the analytic
solution obtained by \citetalias{bz77}, shown by the dotted lines. For larger spins
up to $\astar\lesssim0.9$, both the dimensionless angular field line velocity
$\omega$ and  the
enclosed current $i$ increase with increasing $\astar$ (Figures~\ref{fig4}a,b).
According to equation \eqref{eq:jetpower}, this should result in an increase in
the normalized jet power $p$, as confirmed in Figure~\ref{fig4}c.
This is the reason for the small but systematic increase in jet power
above the estimate~\eqref{eq:bzaeff}.  For $\astar\gtrsim0.95$, we find
that $\omega$, $i$, and $p$ all decrease relative to
\eqref{eq:bzaeff}, causing a flattening of the jet power at these
extreme spins.  The reason for the decreased power is related
to a change in the poloidal field geometry of the jet near the BH
horizon (see \S\ref{sec:thickdisk} and Appendices \ref{app:bz4} and \ref{app:bz6}): while at low spins the magnetic field is nearly uniform
across the jet for all of our models, at high spins the poloidal field
becomes non-uniform with a maximum field strength at the jet axis and
a minimum near the wall.  Since it is the field geometry near the BH
that sets the power output (see footnote~\ref{ftn:locality} and
Appendix~\ref{app:bz2}), it is
logical that these changes in the field geometry lead to changes
in the power output.  We demonstrate this in \S\ref{sec:thickdisk}
(see also Appendices~\ref{app:bz4} and \ref{app:bz6}).

\begin{figure}
  \begin{center}
      \includegraphics[width=\figwidthfactor\columnwidth]{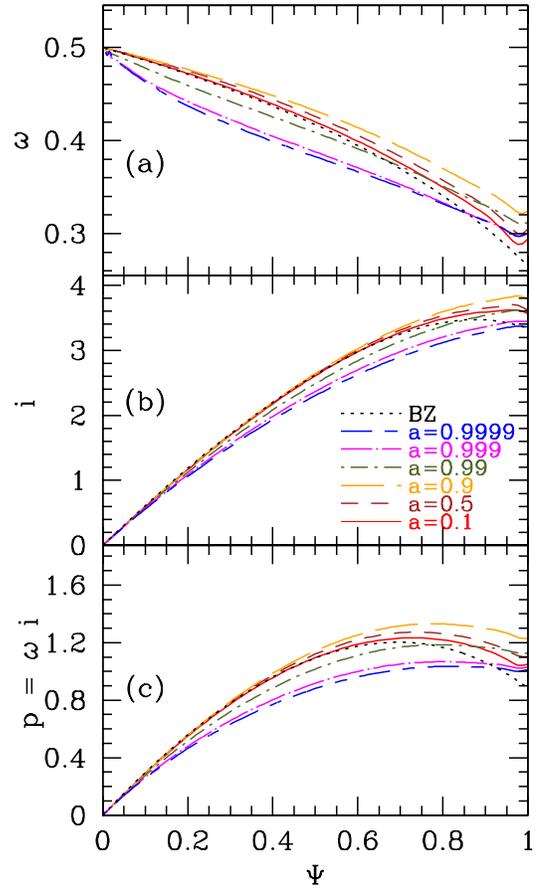}%
  \end{center}
  \caption{
    Similar to Figure~\ref{fig4} but for a paraboloidal jet
    ($\nu=1,r_0=1$). Comparison with Figure~\ref{fig4} shows that, for
    the same BH spin, the angular velocity $\omega$ is smaller and the
    enclosed current $i$ larger in a paraboloidal jet compared
    to a monopolar jet. These two effects
    combine to give a smaller power output $p\equiv\omega{}i$ in the
    paraboloidal solution. Note that, whereas a
    monopolar jet rotates more or less like a rigid body,
    a paraboloidal jet has a significant variation of $\omega$ across
    its cross-section.}
  \label{fig5}
\end{figure}

Figure~\ref{fig5} shows that paraboloidal jets exhibit very similar
trends with increasing spin as their monopolar counterparts.  The
differences are in details, e.g., the angular velocity profile
(Figure~\ref{fig5}c) is now non-uniform even for low spin values, as
predicted by the \citetalias{bz77} analytic solution shown in the
figure with dotted lines.  The agreement with the analytic solution is
not as perfect as for the monopolar model, but this is because the
poloidal field line shape near the BH in our $\nu=1,r_0=1$
paraboloidal jet differs slightly from the \citetalias{bz77}
paraboloidal shape.  For our numerical \citetalias{bz77} paraboloidal
jets the agreement with the analytic solution is very good.

In all our numerical jets the conserved quantities $I(\Psi)$ and
$\Omega(\Psi)$ are preserved along field lines to better than
$10$\%.  We reran a selection of models at twice the fiducial
resolution in both the radial and angular directions.  We found
differences of less than $5$\% in the total power output, indicating
that our models are well-converged.

\begin{figure*}
  \begin{center}
      \includegraphics[width=0.8\textwidth]{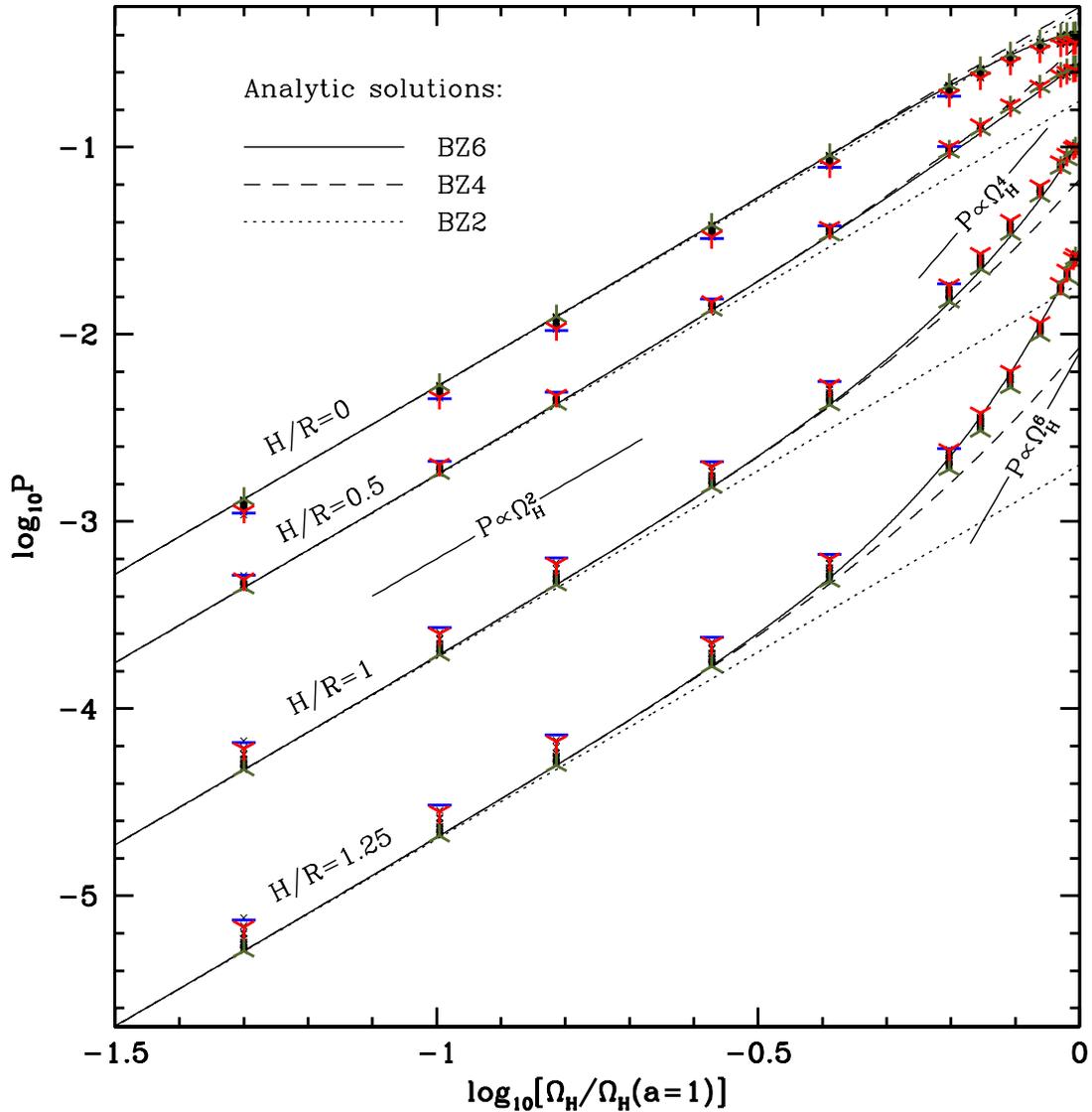}%
  \end{center}
  \caption{Jet power output of various models as a function of
    dimensionless BH horizon spin
    frequency $\omega_{\rm H}=\OmegaH/\OmegaH(a=1)$ (see eq.\ \ref{eq:aeff}) for four different choices of disk thickness:
    razor-thin disk with $H/R=0$, thicker disks with
    $H/R=0.5$, $1$, and $1.25$. For reference, the following 
    values of BH spin, $a=\{0.2,0.5,0.9,1\}$, correspond to approximately
    the following values of $\omega_{\rm H}$: $\{0.1,0.27,0.63,1\}$.
    Numerical results for different models
    are shown with different symbols (see Fig.~\ref{fig3} caption for
    details). The second order analytic BZ2 solution is shown with 
    dotted lines, the $4$th order BZ4 solution with dashed lines, and
    the $6$th order BZ6 solution with solid lines.  As the disk becomes
    thicker, the spin dependence of jet power becomes steeper. This
    steepening is most accurately described by the BZ6 formula.
    Note that this formula also reproduces the flattening of the jet
    power for razor-thin disks ($H/R=0$) at high spins (Figure~\ref{fig3}).  For
    reference, we also show the slopes of various power-law scalings
    $P\propto\OmegaH^n$ with straight line segments.  }
  \label{fig6}
\end{figure*}

\subsection{Power Output of Black Holes with Thick Disks}
\label{sec:thickdisk}

In all the models we described so far, the base of each polar jet covered a
full $2\pi$ steradians at the BH horizon.  However, observational
evidence strongly suggests that low-luminosity BHs ($\lambda<0.01$,
\S\ref{sec_intro}) are surrounded by thick accretion disks or ADAFs
\citep{nm08} with thicknesses $H/R\sim1$. Here and below by the ``disk
thickness'' $H/R$ we mean the angular extent at the BH horizon
of the region exterior to the Poynting-dominated jet, i.e., the total thickness of the
gaseous disk plus any magnetized corona or heavily mass-loaded wind above the disk.
When a BH is surrounded by a thick disk/corona, equatorial field lines
from the BH at lower latitudes pass through the disk/corona, become
turbulent and produce a slow baryon-rich wind, whereas polar field
lines at higher latitudes lie away from the disk gas and produce a
Poynting-dominated relativistic jet \citep{mck05}.  How does this
effect modify the dependence of jet power on BH spin?

Assuming that both the total magnetic flux $\Phi_{\rm tot}$ threading
the BH horizon and the angular thickness $H/R$ of the accretion
disk/corona are
independent of the BH spin, we can compute the power that is
emitted in the Poynting-dominated region of the jet.  We assume that
the models we have described earlier continue to be valid, except that
we integrate the jet power only over field lines that cross the
horizon outside the $\pm H/R$ zone of the disk/corona.
This procedure is well-motivated by general relativistic magnetohydrodynamical
simulations of thick accretion disks which show that the relativistic jet
subtends a well-defined solid angle for a given gas pressure scale height \citep{mck04}.
We consider models with $H/R=0.5$, $1$, and $1.25$,
and compare the results with the case of a razor-thin disk ($H/R=0$).

The results for the jet power as a function of disk thickness and spin
are shown in Figure~\ref{fig6}.  As we have already seen, the jet
power for a razor-thin disk scales as $P\propto\OmegaH^2$ (this
scaling is shown with dotted lines) until $\astar\lesssim0.95$ after
which it levels off. As the disk becomes thicker, the scaling changes
qualitatively.  For all thicknesses, $H/R=0.5$, $1$, and $1.25$, the power
dependence on the spin follows the same $\OmegaH^2$ power-law at low
spins.  However, at higher values of $\astar$, the power increases
more steeply.  The break occurs roughly at $\astar\sim0.7$,
with a moderate dependence on the disk thickness.
Above the break for the case $H/R=1$ we have
$P\propto\OmegaH^{4}$ and for the case $H/R=1.25$ we have
$P\propto\OmegaH^{6}$. This steep dependence is similar to what was
observed by \citet{mck05} in his numerical simulations of turbulent
accreting tori.\footnote{Note that only even powers can enter the expansion of
the jet power in terms of $\OmegaH$ because the power is an even
function of $\OmegaH$.} We observe the same steep power dependence
also in our ideal GRMHD simulations (\S\ref{sec_intro}).

\begin{figure}
  \begin{center}
      \includegraphics[width=\columnwidth]{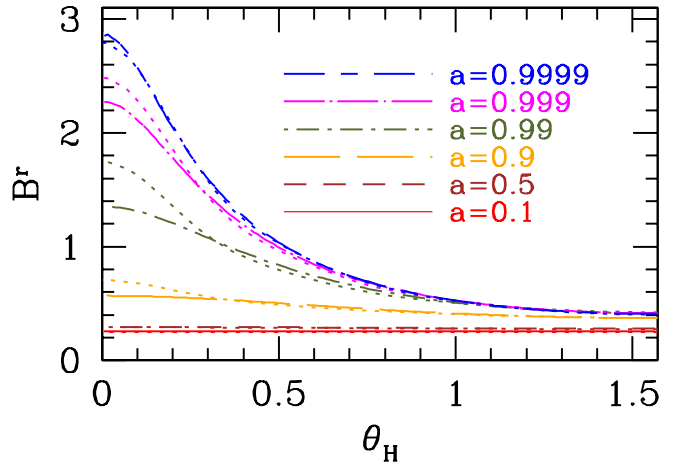}%
  \end{center}
  \caption{Radial contravariant component of the magnetic field
    strength $B^r$ evaluated at the BH horizon in a monopolar model
    ($\nu=0,r_0=0$) as a function of polar angle $\theta_{\rm H}$, for
    different values of BH spin (see legend). As the spin of a BH is
    increased, magnetic field lines progressively bunch up toward the
    BH rotation axis, resulting in an increased magnetic field strength
    close to the axis at small $\theta_{\rm H}$ (this effect can also be seen in
    Figs.~\ref{fig1}--\ref{fig2}). Dotted colored lines show the
    high-order analytic solution, while the various other lines show the
    numerical solution (see legend). }
  \label{fig7}
\end{figure}

We now explain the reason for the steep dependence of jet power on
$\OmegaH$ as $\astar\to1$.  We saw in Figures~\ref{fig1} and
\ref{fig2} that as the BH spin increases, magnetic field lines
rearrange laterally and concentrate around the axis of rotation (see
\citealt{km07} for an explanation of this effect in terms of hoop
stresses).  Figure~\ref{fig7} shows that this leads to a non-uniform
distribution of radial magnetic field $B^r$ across the jet, with $B^r$
having a maximum near the rotation axis and a minimum near the jet
boundary.  Since the electromagnetic energy flux of a BH is
proportional to $(B^r)^2$ at the BH horizon (see
equation~\ref{eq:bhpowerapp}), the concentration of magnetic flux
near the rotation axis leads to a progressively larger fraction
of the total energy output of the BH to be emitted in the polar region, giving a
steeper dependence of jet power on the BH spin in the presence
of a thick disk ($H/R\sim1$).  In a
related context, \citet{macdonald1984} and \citet{km07} have studied
how magnetic flux is pulled in by
a spinning BH by considering magnetic hoop stresses.
\citet{macdonald1984} appears to have missed the strength of this effect
by mostly investigating models with relatively small $\astar\lesssim 0.7$
and by primarily looking for a change in the total magnetic flux accumulated at the horizon
rather than measuring the flux redistribution on the horizon.

We now describe an analytic approach for understanding 
the steep dependence of jet power on  BH spin for thick disks with $H/R\sim1$.  
While the split-monopolar magnetic field is an exact
solution for non-spinning BHs, for spinning BHs the dragging of frames
induces a spin-dependent perturbation to the split-monopolar magnetic
field geometry.  It is this perturbation that causes field lines to
move preferentially toward the rotational axis.  By performing an
expansion in powers of $a$, \citetalias{bz77} determined this
perturbation of the magnetic field geometry to the lowest (second)
order in BH spin $a$ (see also \citealt{mck04} and \citealt{tn08}).
Accounting for this perturbation in the field geometry, \citet{tn08}
determined the BH power output more accurately than the original
\citetalias{bz77} derivation, to the $4$th order in BH spin $a$.  As
we noted in \S\ref{sec:thindisk}, expansions in powers of BH frequency
$\OmegaH$ are more accurate at high spins than expansions in powers of
the BH spin $a$.  Therefore, in Appendix~\ref{app:bz4} we perform an
equivalent expansion in terms of $\OmegaH$.\footnote{This expansion
  reduces to the expansion~\eqref{eq:bz4} in the limit of $H/R\to0$.}
Figure~\ref{fig6} shows with dashed lines the power output of our
$\OmegaH$-based model, which we refer to as BZ4. Clearly, it provides a more accurate
approximation for the power of our numerical jets than the $2$nd order
accurate BZ2 solution shown with the dotted lines.
However, as we saw in \S\ref{sec:thindisk} and as
is clear also from Figure~\ref{fig6}, the BZ4 solution still does not
capture a few important effects: (i) for BHs with razor-thin disks, it
does not capture the flattening of the power output at $a\gtrsim0.95$
and thereby over-predicts the numerical results by as much as $25$\%,
and (ii) for thick disks this solution under-predicts the numerical
power by as much as $70$\%.

To improve our analytic model, we have used the results of our numerical simulations
to determine higher-order corrections to both the field
geometry and the power of the BH at high spins.  Firstly, we have obtained a higher-order
accurate numerically-motivated expansion in powers of $\OmegaH$
for the \emph{magnetic field} $B^r$
at the BH horizon (Appendix~\ref{app:bz6}
provides the details).  Shown with dotted lines in Figure~\ref{fig7} for a
wide range of $\astar$, this analytic approximation for $B^r$ accurately reproduces the
numerical angular profile of magnetic field
for a wide range of polar angles, $\theta_{\rm H}\gtrsim0.3$ or
$H/R\lesssim1.3$. Secondly, based on this higher-order magnetic field profile, we have
analytically computed the $6$th order accurate approximation for the
\emph{BH power output}, which we refer to as model BZ6 (see Appendix~\ref{app:bz6}
for more detail).  Figure~\ref{fig6} shows
the results with the solid lines.  Not only does the BZ6 expansion
correctly capture the flattening of the BH power at high BH spins for
razor-thin disks, it also provides a
significantly more accurate approximation to the jet power output for
thicker disks. For instance, for a thick disk with $H/R=1.25$, the
BZ6 expansion is about a factor of $20$ more accurate
than the BZ4 expansion.

\section{Discussion}
\label{sec:discussion}

Figure \ref{fig3} shows that, regardless of the geometry of the confining wall, the total power
output of a magnetized relativistic spinning BH with a razor-thin disk varies
as $P = k\Phi_{\rm tot}^2
\OmegaH^2$ (equation~\ref{eq:bzaeff}), where $\Phi_{\rm tot}$ is the
total magnetic flux threading the BH horizon,
$\OmegaH=a/(2r_{\rm H})$ is the BH horizon frequency, and $r_{\rm
  H}=M\left[1+(1-\astar^2)^{1/2}\right]$ is the radius of the BH
horizon in units of $G/c^2$.  This modified BZ2 scaling
is slightly steeper than the original \citetalias{bz77} scaling $P =
k\Phi_{\rm tot}^2(\astar/4M)^2$ (equation~\ref{eq:bzori}).  A
more accurate BZ6 scaling (equation~\ref{eq:bzbf})
accurately reproduces the power-spin dependence for all values of BH
spin $a$, including the limit $\astar\to1$.

In the context of the radio loud/quiet dichotomy of AGN, following
\citet{ssl07} let us assume that supermassive BHs in elliptical
galaxies, which manifest themselves as radio loud AGN, have higher
spin parameters with a median $\astar\sim0.9$, while the BHs in
spirals, the radio quiet AGN, have a lower median $\astar\sim0.3$ (e.g.,
\citealt{volonteri07}).  Figure~\ref{fig3} then suggests that the jet
powers in the two classes of objects (assuming similar values of
$\Phi_{\rm tot}$) would differ by a factor $\sim20$.  However,
radio loud AGN and radio quiet AGN differ in their radio luminosities
by a factor $\sim10^3$ \citep{ssl07}.  What could be the reason for
such a large dichotomy in jet power?

One motivation for the present study was to investigate whether there
is any strong non-linearity in BH physics that might cause the jet
power to increase very rapidly as the BH spin approaches unity.  If
this were the case, one could pursue a scenario in which radio loud
AGN are associated with nearly extremal Kerr BHs.  Unfortunately, our
numerical results indicate that non-linearity hardly helps.  Because
the total BH power output scales as $\OmegaH^2$ (equation~\ref{eq:aeff}) rather than
simply as $\astar^2$, there is a slightly steeper increase of power with
$\astar$ as the spin approaches unity.  However, the scaling actually
becomes shallower once $\astar$ increases above $\sim0.99$.  We have
carefully checked the convergence of our models and we are confident
that the results are not affected by numerical errors.  Therefore,
there is not much room for increasing the power of radio loud AGN jets
by pushing $\astar$ arbitrarily close to unity.

Therefore, since there is not much wiggle room at the radio loud end,
we need to postulate that radio quiet AGN have very low values of $\astar$,
say $\astar\sim0.03$.  This is uncomfortably low.  It is certainly
feasible for an occasional BH to be spinning so slowly, but to have an
entire population of BHs (radio quiet sources in spirals) with a
median $\astar$ of order $0.03$ seems far-fetched.  It would require spiral
galaxies not to have experienced any significant mergers.
Furthermore, the BHs in their nuclei should have accreted mass entirely through
minor mergers with smaller companions, each with a tiny mass and
with a random orientation of angular momentum
\citep{hughesblandford03,gammie_bh_spin_evolution_2004,bv08}.

The second motivation for the present study was to investigate if the
geometry of the confining funnel along which the jet propagates, which
may be different for rapidly-spinning and slowly-spinning BHs, could
lead to a substantial change in the jet power.  This too turns out not
to be the case, within the context of razor-thin disks.  We have tried a wide range of geometries for the jet,
as described in \S\ref{sec:num} (see also Figs. \ref{fig1}, \ref{fig2}), but the jet
power varies by no more than $20\%$ for a fixed BH spin and magnetic
flux.

Our third motivation was to investigate the correctness
of the results of \citet{mck05} who obtained a steeper dependence in the jet power
compared to the total BH power.
For this purpose, we considered in
\S\ref{sec:thickdisk} an additional geometrical effect,  viz., varying the solid angle subtended by the base of
the jet.  Such a variation is expected if the accretion disk is geometrically thick.
As an example, let us consider the results corresponding to a BH
surrounded by a disk of angular extent $H/R=1$,
and let us further assume that radio loud AGN have spin parameters
very close to unity and radio quiet AGN have more modest values of $a$.  Then Figure~\ref{fig6} shows that it is possible
to explain the radio loud/quiet dichotomy, i.e., a factor of $10^3$ in
jet power, if radio loud AGN have $\astar\to1$ and radio quiet systems
have $\astar\sim0.15$.  This is a lot more comfortable than the
requirement $\astar\sim0.03$ that we found earlier for a razor-thin disk.  Indeed, if the
disk/corona is even thicker than $H/R\sim1$, which is not 
unreasonable\footnote{We note that $H/R$ in this context refers to all the
gas-dominated regions of the flow: the accretion disk proper, the corona and
the disk wind.  The net half-angle subtended by all these components could
equal a radian or more in the case of a thick accretion flow.}, the effect is even
stronger (see Figure~\ref{fig6}), and the radio loud/quiet dichotomy
can be explained with quite modest changes in spin.

There are several other effects that we did not consider
which might either enhance or diminish the effect of rapid spin.
For instance, we assumed that the total magnetic flux threading the horizon
is a constant, independent of BH spin.
Any mechanism that enhances the total magnetic flux threading the horizon
would lead to a larger BH power output~\citep{mck05,hk06,km07,rgb06,gar09}.
For example, \citet{mck05} found that the magnetic flux across the entire horizon
scales as $\Phi_{\rm{}tot}\propto \OmegaH^{1/2}$,
so the total power will increase by another factor of $\Phi_{\rm{}tot}^2\propto \OmegaH$,
which further diminishes the range of BH spins
required to explain the radio loud/quiet dichotomy.
At first sight, it would appear that $\Phi_{\rm tot}$ is determined by conditions far
from the BH, e.g., the net magnetic flux of the gas supplied to the
accretion disk on the outside, and the ability of the disk gas to
transport this field in.
However, once the field has been transported to the center,
two general relativistic effects kick in.
First, the spin of the BH determines the size of the plunging region,
and larger plunging regions can accumulate more flux \citep{gar09}.
Note that magnetic flux can also be transported
to the BH through the corona outside the accretion disk~\citep{rl08,bhk09}.
Second, the frame-dragging of space-time near the BH within the
ergospheric region leads to currents that generate hoop stresses
pulling magnetic flux toward the horizon \citep{km07}.
These two effects are non-trivially coupled,
although for prograde BH spins the hoop stresses appear to dominate \citep{mck05},
while for retrograde BH spins the size of the plunging region may dominate \citep{gar09}.
Another possibility is that a stronger dependence on spin could occur
if some field lines attach between the disk and the BH \citep{yewang05},
although such configurations are not seen
in general relativistic magnetohydrodynamical simulations of accretion disks \citep{hirose04,mck05}.

Yet another possibility is that mass-loading of AGN jets
might have a large effect on the jet power.
For example, the \citetalias{bz77} mechanism
only operates at sufficiently high magnetization for a given BH spin.
Thus, the magnetization and BH spin can together introduce a ``magnetic switch''
mechanism that can trigger powerful jet formation
from the BH~\citep{takahashi1990,meier97,meier99,kb09}.
Another type of magnetic switch can be due to changes in the field geometry
from dipolar to multipolar, which leads to significant mass-loading of the polar
regions \citep{mck04,bhk07,mb09} and a factor of $\sim 10$ weaker total BH power output.
The physics of jet mass-loading is presently uncertain,
and this question needs to be investigated in more detail.
Finally, changes in the disk thickness may also result in differences in the amount
of magnetic flux accumulated or generated by turbulence near the BH \citep{meier01}.

Finally, we note that we have implicitly assumed in this paper that
the radio luminosity of a jet is directly proportional to the total
energy flux (Poynting and kinetic power) carried by the jet.  Perhaps
this is not the case.  Any non-linearity in the mapping between
radiative luminosity and jet power could have important consequences.
In particular, we note that the interstellar medium (ISM) in a typical
elliptical galaxy is very different from that in a typical spiral.
Since the radio emission in a jet is produced when the jet interacts
with the external ISM, this difference may well lead to a strong
effect on the radio loudness of the jet.

In application to gamma-ray bursts (GRBs) and collapsars, an
interesting question is whether they are powered by the BZ mechanism
through an outflow from a central BH or
by an outflow from an accretion disk.  \citet{kb09} suggest that the BZ
effect is operating in such a scenario (however, see \citealt{nagataki09}).

\section{Conclusions}
\label{sec:conclusions}

We set out in this paper to explain the radio loud/quiet dichotomy of AGN
in the context of the BH spin paradigm.
For razor-thin disks, we found that BH spin alone is insufficient to explain the observations
even if the radio loud and radio quiet populations have very different merger and accretion histories.
However, we found that the presence of a thick disk, such as an ADAF,
can significantly enhance the spin dependence of the power output,
to the extent that it can reasonably account for the observed radio loud/quiet dichotomy.
Our only modification to the revised spin paradigm of \citet{ssl07}
is that both populations should contain a BH surrounded by a thick disk such
that the jet subtends  a small solid angle around the polar axis.

These results were obtained by performing general relativistic
numerical simulations of collimated force-free and MHD jets
from spinning magnetized BHs for a wide range of spins (up to $a=0.9999$) and jet confinement
geometries.  We showed that, regardless of the geometry, a BH
threaded with a magnetic flux $\Phi_{\rm tot}$ and surrounded by a razor-thin disk produces
a jet with power $P\approx k\Phi_{\rm tot}^2\OmegaH^2$, where $k$ is a known
constant factor which depends only weakly on the field geometry, $r_{\rm H} =
M[1+(1-a^2)^{1/2}]$ is the radius of the BH horizon, and $\OmegaH=a/(2r_{\rm H})$
is the angular frequency of the BH.
This result gives a somewhat steeper dependence of jet power on $\astar$
compared to the original scaling $P\propto \astar^2$ 
obtained
by \citetalias{bz77}.  Nevertheless, we conclude that for a fixed magnetic flux $\Phi_{\rm tot}$, even this revised scaling is much too shallow to
explain the radio loud/quiet dichotomy of AGN.  Our goal, therefore, was to 
identify 
any other effect that may cause the jet power to
depend more steeply on BH spin.

We found that such an effect naturally exists.  We showed that the
power output of a BH surrounded by a thick accretion disk with
$H/R\sim1$ (this is the effective thickness of the disk, corona and
mass-loaded disk wind), as expected in systems with
advection-dominated accretion flows (ADAFs, \citealt{nm08}), is $P\propto
\OmegaH^4$, and even $\propto\OmegaH^6$ for very thick disks
(\S\ref{sec:thickdisk}).  In this case we can explain the radio
loud/quiet dichotomy by having two different populations of galaxies
with modestly different BH spins.  For the case $H/R=1$, the radio
loud population needs to have large spins $a\simeq1$ while
the radio quiet AGN population needs to have $a\simeq0.15$.
Such spin values may plausibly result from differences in the
merger and accretion histories of supermassive BHs in elliptical
and spiral galaxies (\S\ref{sec:discussion}).  

We worked out in the Appendices a first principles analytic model
which accurately reproduces our numerical results for the jet power over a
wide range of BH spin and disk thickness (Figures~\ref{fig3}, \ref{fig6}).

\acknowledgements

We thank Vasily Beskin and Serguei Komissarov for useful comments on
the manuscript.  This work was supported in part by NASA grant
NNX08AH32G (AT \& RN), NSF grant AST-0805832 (AT \& RN),
NASA Chandra Fellowship PF7-80048 (JCM), and by
NSF through TeraGrid resources~\citep{catlett2007tao}
provided by the Louisiana Optical Network Initiative
(\href{http://www.loni.org}{http://www.loni.org}).

\appendix

\section{Second-order--accurate Expansion of Black Hole Power}
\label{app:bz2}

In this section we present a compact derivation of the
\citetalias{bz77} effect.  We determine the power output of a
spinning BH embedded into an externally-imposed split-monopolar
magnetic field.  This magnetic field is given by the flux
function~\eqref{eq:psi} with $\nu = 0$ and $r_0=0$.  The main
difference of this derivation is that we perform it in the powers of
the ``natural'' variable -- the BH angular frequency $\OmegaH$ that
plays an important role in determining the BH power output.  The power
output density evaluated at the horizon of a spinning BH is
(\citealt{bz77}; \citealt{mck04})
\begin{equation}
  \label{eq:bhpowerapp}
  F_{\rm E}(\theta)  = \left[2 (B^r)^2 \Omega (\OmegaH-\Omega) r M\sin^2\theta\right]\Bigr|_{r=r_{\rm H}},
\end{equation}
where the quantity $\Omega$ is the angular frequency of magnetic
field lines at the BH horizon and $B^r$ is the radial field strength
at the horizon.  

In order to determine the power output~\eqref{eq:bhpowerapp}, we need
to know two quantities as functions of polar angle at the BH horizon:
the radial magnetic field, $B^r$, and the field line angular
frequency, $\Omega$.  The element of the magnetic flux $d\Phi$ through
the BH horizon is related to the radial magnetic field at the horizon
through the following differential,
\begin{equation}
  \label{eq:bdef}
  d\Phi = 2\pi d\Psi=2\pi B^r \gdet\,d\theta,
\end{equation}
where $g$ is the
determinant of the Kerr metric in the Boyer-Lindquist
coordinates, $\gdet=(r^2+a^2\cos^2\theta)\sin\theta$. This formula closely resembles its cousin in the
spherical polar coordinates and flat space where one has
$\gdet=r^2\sin\theta$.  More generally, in place of $B^r$ we can use
any vector field (e.g., energy flux), and the result is the differential of
that flux. Assuming that the perturbations to the magnetic
field away from a perfect split-monopole are higher order, we neglect them
and obtain, by differencing the flux function \eqref{eq:psi} (with
$\nu=0$ and $r_0=0$) according to \eqref{eq:bdef}, an identical result
to that in flat space:
\begin{equation}
  \label{eq:brmono}
  B^r=\Psi_{\rm tot}/r^2,
\end{equation}
where we neglected terms of order $\OmegaH^2$ and higher.  While the
distribution of $\Omega$ needs to be self-consistently determined by
solving the non-linear equations describing the balance of
electromagnetic fields (we do so numerically in \S\ref{sec:results}),
here we make a simple yet accurate estimate based on the energy
argument. Let us assume that the system chooses such a distribution
of $\Omega$ that it causes an extremum in BH power
output~\eqref{eq:bhpowerapp}.  Such a value is clearly
\begin{equation}
  \label{eq:omegamono}
  \Omega = \OmegaH/2
\end{equation}
since it maximizes the BH power output~\eqref{eq:bhpowerapp} \citep[see,
e.g.,][]{bk00}.  Despite the simplicity of this estimate, it is
remarkably close to the true solution for the split-monopolar geometry
as obtained from the numerical simulations (\S\ref{sec:results}).
Plugging equations \eqref{eq:brmono} and \eqref{eq:omegamono} into the
power output density \eqref{eq:bhpowerapp}, we obtain
\begin{equation}
  \label{eq:fesimple}
  F_{\rm E}(\theta) = 2\left(\Psi_{\rm tot}r^{-2}\right)^2 (\OmegaH/2)^2 r
  M\sin^2\theta.
\end{equation}
Integrating up this power output density in angle in the same way as
we integrated the magnetic field in equation~\eqref{eq:bdef}, we
obtain the full power output into jets with an opening angle $\theta_j$:
\begin{equation}
  \label{eq:powertheta}
  P = 2\times2\pi\int_0^{\theta_j} F_{\rm E}(\theta)  \gdet\Bigr|_{r=r_{\rm H}}d\theta,
\end{equation}
where the extra factor of $2$ accounts for the fact that there are two
jets, one in the northern
and one in the southern hemisphere.
Note that we are interested in an expansion of power up to 2nd order
in $\OmegaH$. Since the factor $F_{\rm E}(\theta)\propto \OmegaH^2$ is
already second order in $\OmegaH$ (equation~\ref{eq:fesimple}), we can
without loss of accuracy evaluate the formula~\eqref{eq:powertheta} at
$r=r_{\rm H}(a=0)=2M$ and replace $\gdet$ with
$\left[(2M)^2\sin\theta\right]$.  After plugging
into~\eqref{eq:powertheta} for $F_{\rm E}(\theta)$
using~\eqref{eq:fesimple} and evaluating the integral out to
$\theta_j=\pi/2$, i.e., computing the full power output of the BH,
we get:
\begin{equation}
  \label{eq:powerthetasimple}
  P = \pi\Psi_{\rm tot}^2 \OmegaH^2
  \int_0^{\pi/2}\sin^3\theta\,d\theta =
  2\pi\Psi_{\rm tot}^2\OmegaH^2/3,
\end{equation}
which is accurate to  second order in $\OmegaH$.
In terms of the magnetic flux $\Phi_{\rm tot}=2\pi\Psi_{\rm tot}$, the
formula becomes
\begin{equation}
  \label{eq:powerthetasimple2}
  P = k \Phi_{\rm tot}^2\OmegaH^2,
\end{equation}
with $k=1/(6\pi)$, which reproduces \eqref{eq:bzaeff}.

\section{Fourth-order--accurate Expansion of Black Hole Power}
\label{app:bz4}

In Appendix \ref{app:bz2} we have derived a second order accurate
expression for power output of the BH in terms of the hole frequency
$\OmegaH$.  The BH was embedded with a split-monopolar magnetic field.
Let us now improve the accuracy of the previous derivation, this time
retaining the higher order terms, up to $\OmegaH^4$. A similar
derivation was performed by \citet{tn08} but in powers of BH spin
$a$.  Here we derive the expansion in terms of the natural
variable $\OmegaH$ which allows to use the expansion for nearly
maximally spinning BHs. We also explicitly present the angular
dependence of the BH power output.

In order to obtain a higher-order approximation to the power, this
time we need to keep some of the higher terms we neglected in
equations for $B^r$ \eqref{eq:brmono} and $\Omega$
\eqref{eq:omegamono}.  Since $\Omega$ is an odd function of BH
frequency, it contains only odd powers of $\OmegaH$, therefore a
higher order approximation for it has the following form:
\begin{equation}
  \label{eq:omega3rd}
  \Omega=\myfrac{1}{2}\OmegaH+\mathcal O(\OmegaH^3),
\end{equation}
where $\mathcal O(\OmegaH^3)$ denotes any third order or higher order
terms in $\OmegaH$. Since $P\propto\OmegaH^2$, these higher-order
terms contribute to the power output only terms of order higher than
$\mathcal O(\OmegaH^4)$, therefore we neglect them. We do need,
however,  to include the terms that come
from the higher order expansion of $B^r$.  \citetalias{bz77} showed
that the dragging of frames by the spinning BH perturbs the magnetic
field away from an exact split-monopole and have derived a second
order correction to the flux function, $\Psi_2$, in powers of BH spin
so that the full flux function has the form
\begin{eqnarray}
  \Psi(r,\theta) &=& \Psi_0(\theta) + a^2 \Psi_2(r,\theta) +\mathcal O(a^4) \\
  &=& \Psi_0(\theta) + 16 \OmegaH^2 \Psi_2(r,\theta) + \mathcal
  O(\OmegaH^4),
  \label{eq:app:bzpsi}
\end{eqnarray}
where we have used $a = 4\OmegaH + \mathcal O(\OmegaH^3)$.  Here
the zeroth and second order perturbations to the flux function are
\begin{equation}
  \label{eq:app:bzpsi1}
  \Psi_0(\theta) = 1 - \cos\theta, \quad \Psi_2(r,\theta) = f(r) \sin^2\theta\cos\theta,
\end{equation}
where $f(r)$ is a known function of radius $r$, but for further
discussion only its value at the horizon of a non-spinning BH, $f(r=2) =
\left(56-3\pi^2\right)/45$, is needed (this is because in the
expansion \eqref{eq:app:bzpsi}  we formally
evaluate the coefficients at  $\OmegaH=0$, $r=r_{\rm H}=2$).

Combining expressions \eqref{eq:app:bzpsi}, \eqref{eq:app:bzpsi1} and
\eqref{eq:bdef}, we obtain the $2$nd-order--accurate radial magnetic
field at the BH horizon:
\begin{equation}
  \label{eq:brbz4}
  B^r = \frac{\left(1+4 {\OmegaH}^2\right)^2 \left[9+{\OmegaH}^2 (-49+6 \pi ^2)(1+3 \cos2\theta)\right]}{9 \left(r_{\rm H}^2\left(1+4 {\OmegaH}^2\right)^2+16 {\OmegaH}^2 \cos\theta^2\right)}.
\end{equation}
Combining this expression with \eqref{eq:bhpowerapp},
\eqref{eq:omegamono} and plugging into \eqref{eq:powertheta}, we
numerically obtain the angle-dependent enclosed power $P^{\rm BZ4}(\theta,\OmegaH)$ shown in
Figure~\ref{fig6} as $P^{\rm BZ4}(\theta=\pi/2-H/R,\OmegaH)$ with dashed lines.  This result, expanded to $4$th
order in powers of $\OmegaH$, is:
\begin{eqnarray}
  P^{\rm BZ4}(\theta,\OmegaH)&\approx& \pi \OmegaH^2  \left[\myfrac{4}{3} \sin^4(\theta/2) (\cos \theta +2)\right]\notag\\
   &+&\pi  \OmegaH^4 \bigl[90 \left(3 \pi ^2-32\right) \cos
   \theta +\left(970-105 \pi^2\right) 
   \cos 3 \theta 
\ifthenelse{\equal{\useiop}{-1}}{
  \notag\\
  &\phantom{=}&\phantom{\pi  \OmegaH^4 \bigl[}
}
   {}+9\left(3 \pi^2-26\right) \cos 5 \theta+32 \left(67-6 \pi ^2\right)\bigr]/{270}
   + \mathcal O(\OmegaH^6)
  \label{eq:bz4thpwr}
\end{eqnarray}
Clearly, at low spins, the second-order piece dominates, which we show
in Figure \ref{fig6} with dotted lines.  However, at high spins, the
fourth order piece can become dominant, which is confirmed in
Figure~\ref{fig6}.  To see this more clearly, we perform an expansion
of~\eqref{eq:bz4thpwr} in powers of disk/corona thickness,
$H/R\equiv\pi/2-\theta$, and obtain to second order in $H/R$:
\begin{equation}
  \label{eq:bz4hor}
  P^{\rm BZ4}(H/R) \approx\Omega _H^2 \left\{2.09-3.14 H/R + \mathcal O[(H/R)^3]\right\}+\Omega_H^4
  \left\{2.9+1.7 H/R + \mathcal O[(H/R)^3]\right\} + \mathcal O(\OmegaH^6),
\end{equation}
where for clarity we have numerically evaluated the coefficients to
two decimal places.  This expansion makes it clear that as $H/R$
increases, the relative importance of the fourth order term increases.
This also explains why in Figure~\ref{fig6} the
$P\propto\OmegaH^4$ dependence becomes more prominent for larger values
of $H/R$ as opposed to smaller values.

Finally, we note that at the midplane the power output takes the
following form (expanded up to $4$th order in $\OmegaH$):
\begin{equation}
  \label{eq:bz4pwrequator}
  P^{\rm BZ4}(\theta=\pi/2)\approx\myfrac{2\pi}{3}\Psi_{\rm
    tot}^2\left[\Omega _H^2+\Omega _H^4 \,8 \left(67-6
      \pi^2\right)/45\right]+\mathcal O(\OmegaH^6)=
   \myfrac{2\pi}{3}\Psi_{\rm tot}^2\left[\Omega _H^2+\alpha\Omega_H^4\right],
\end{equation}
where $\alpha = 8 \left(67-6 \pi^2\right)/45 \approx 1.38$.  In this
formula we have
reintroduced $\Psi_{\rm tot}$ which was set to unity for the rest of
this section.  This result can also be expressed in terms of the total
flux in the jet using $\Phi_{\rm tot}=2\pi\Psi_{\rm tot}$.

\section{Sixth-order--Accurate Expansion of Black Hole Power}
\label{app:bz6}
Figure \ref{fig6} shows that the fourth-order BZ4 solution
for power is more accurate than the second-order solution. However, at
high BH spin, $a\gtrsim 0.95$, it requires a more than a factor of $3$ correction in
order to reproduce the numerical solution. Also, the fourth
order BZ4 solution does not reproduce the flattening of the power
dependence on the BH spin for razor-thin disks ($H/R=0$) at
$a\gtrsim0.95$.

Inspired by the success of the previous section, we would like to
derive a sixth-order--accurate expression for the power.  However, for
that we would need to know the expansion of the flux function to the
fourth order and of the field angular frequency to the third order.
None of these are known analytically, therefore,
we adopt a numerical approach.

Figure~\ref{fig4}a shows the angular profiles of the field line
rotation frequency $\Omega$ for different BH spins.  While the
deviations from the zeroth order approximation $\Omega=\OmegaH/2$ are
present, their relative magnitude is very small, $\lesssim10$\%.
These $10$\% changes in $\Omega/\OmegaH$ translate into at most $1$\%
changes in the power output because the power depends quadratically on
the magnitude of the higher order correction $(\Omega-\OmegaH/2)$ (see
equation~\ref{eq:bhpower}).  We are interested in the corrections of
order $\sim10{-}70$\% (the level of inaccuracy of the BZ4 solution),
therefore we neglect the higher-order corrections to $\OmegaH$ in
deriving the sixth order solution.

The corrections to the magnetic field shape are, however, dramatic.
Figure~\ref{fig7} shows the angular distribution of the radial
magnetic field $B^r$ at the BH horizon.  As the BH spin increases,
$B^r$ develops a progressively large non-uniformity in angle and
deviates from the $4$th order solution by factors of a few.  We
therefore, attempt to find a numerical fit to the angular magnetic
field dependence at the BH horizon for $a=0.9999$ by fitting to it the
following trial function:
\begin{equation}
  \label{eq:bz6psi}
  \Psi=\Psi_0+16\OmegaH^2\Psi_2+\OmegaH^4\Psi_4,
\end{equation}
where the first two terms are given by equations \eqref{eq:app:bzpsi1}.
We look for the spin-independent part of the third term,
$\Psi_4$, in the following form
\begin{equation}
  \label{eq:bz6psi4guess}
  \Psi_4(\theta)=\sin^2(\theta)
          \left[
             c_1\cos^{\alpha_1}\theta+
             c_2\cos^{\alpha_2}\theta+
             c_3\cos^{\alpha_3}\theta+
             c_4\cos^{\alpha_4}\theta\right],
\end{equation}
where we choose  $\alpha_1=25$, $\alpha_2=7$, $\alpha_3=3$,
$\alpha_4=1$.
In order to determine $4$ coefficients $c_1$--$c_4$, we match the
numerical solution for $B^r$ at $a=0.9999$, shown in
Figure~\ref{fig7}, at $4$ angles: $\theta=0.1$, $0.5$, $0.7$, $\pi/2$.
We find $c_1\approx26.16$, $c_2\approx22.72$, $c_3\approx13.54$,
$c_4\approx2.08$.  Figure~\ref{fig7} shows as dotted colored lines the
solutions due to \eqref{eq:bz6psi} and \eqref{eq:bz6psi4guess}, with
the above values of expansion coefficients, for various values of BH
spin.  Clearly, the analytic fit is a very good match to the power
output at $\theta_{\rm H}\gtrsim0.3$.  Very close to the rotation
axis, however, (at angles smaller than $0.3$) the
fourth-order--accurate solution to the flux function \eqref{eq:bz6psi}
is not enough: while we have a nearly perfect match between our fit
and the numeric solution for $B6$ at high ($a=0.9999$) and mid-range
($a=0.5$) spins, our fit to $B^r$ deviates by up to $\sim25$\% at the
in-between spins ($a\simeq0.9{-}0.99$).  However, the total power
emitted into the range of polar angles $\theta\lesssim0.3$ is very
small, therefore these deviations of our fit from the numerical
solution hardly influence the fit to the power output.

We also considered a direct fit to the vector spherical harmonic
functions that form a complete set for the vector potential
as given by equation B8 in \citet{mck07a}.
However, even an expansion up to $l=10$ did not fit the very steep
behavior of $B^r$ near the polar axis.  However, otherwise,
even only using up to $l=6$ does a reasonable job at fitting
the numerical solution for the total power vs. spin and angle.
This fact and the fact that a power of $25$ for $\cos\theta$
was required to fit the numerical results
demonstrates that the numerical solution at $a\sim 1$ is highly non-linear
with respect to $\theta$ and would be quite difficult to derive analytically.
This proves the usefulness of the numerical simulations.

Now we are in a position to analytically compute the high-order--accurate power of
our jets.
Using~\eqref{eq:bdef},  we compute the radial field on the horizon, $B^r$,
from the fourth-order flux function \eqref{eq:bz6psi}.  We then insert this
field and the field angular frequency $\Omega$ \eqref{eq:omegamono}
into formula~\eqref{eq:powertheta} and obtain the angular-dependent
jet power.
This power, which we refer to as the sixth-order
analytic BZ6 solution, is shown in Figure~\ref{fig6} with solid lines
for  various disk thicknesses and spins.  (We compute these
lines by numerically integrating the analytically-determined power in our jets. 
We note while formally this solution is $6$th order-accurate,
its expansion in powers of $\OmegaH$ contains important terms up 
to $10$th order. This highlights the non-linearity of the problem.)  Clearly,
these lines approximate the numerical data points very well, within
$20$\% for the whole range of disk thicknesses and spins that we have
explored.  Over most of the parameter space the errors are smaller
than this value (they are largest for the thicker disks with
$H/R\gtrsim1.25$ that have the BH spin in the range $0.8\lesssim
a\lesssim0.99$).




\label{lastpage}
\end{document}